\documentclass[8.5pt,twoside,twocolumn,notitlepage]{article}

\usepackage[super,sort&compress,comma]{natbib} 
\usepackage[version=3]{mhchem}
\usepackage{times}
\usepackage{sectsty}
\usepackage{balance} 
\usepackage{graphicx} 
\usepackage{lastpage}
\usepackage[format=plain,singlelinecheck=false,font=small,labelfont=bf,labelsep=space]{caption} 
\usepackage{fancyhdr}
\usepackage{authblk} 
\usepackage{amssymb,amsmath}
\title{Stable Small Bubble Clusters in Two-dimensional Foams}
\author[e]{Kai Zhang}
\author[a]{Chin-Chang Kuo}
\author[a]{Nathaniel See}
\author[b,c,d]{Corey O'Hern}
\author[a]{Michael Dennin \thanks{mdennin@uci.edu}} 

\affil[a]{Department of Physics and Astronomy,  University of California, Irvine 92697, USA }
\affil[b]{Department of Mechanical Engineering and Materials Science, Yale University, New Haven, Connecticut 06520, USA}
\affil[c]{Department of Applied Physics, Yale University, New Haven, Connecticut 06520, USA}
\affil[d]{Department of Physics, Yale University, New Haven, Connecticut 06520, USA}
\affil[e]{Department of Chemical Engineering, Columbia University, New York, New York 10027, USA}

\begin{document}

\maketitle 

\twocolumn[
  \begin{@twocolumnfalse}
\noindent\LARGE{\textbf{Stable Small Bubble Clusters in Two-dimensional Foams}}
\vspace{0.6cm}

\noindent\large{\textbf{Kai Zhang,\textit{$^{e}$}, 
Chin-Chang Kuo\textit{$^{a}$}, Nathaniel See\textit{$^{a}$}, Corey O'Hern,\textit{$^{b,c,d}$}, and Michael Dennin,$^{\ast}$\textit{$^{a}$} }}\vspace{0.5cm}

\noindent\textit{\small{\textit{$^{a}$}\textbf{Department of Physics and Astronomy,  University of California, Irvine 92697, USA}}}\newline
\noindent\textit{\small{\textit{$^{b}$}\textbf{Department of Mechanical Engineering \& Materials Science, Yale University, New Haven, Connecticut 06520, USA}}}\newline
\noindent\textit{\small{\textit{$^{c}$}\textbf{Department of Applied Physics, Yale University, New Haven, Connecticut 06520, USA}}}\newline
\noindent\textit{\small{\textit{$^{d}$}\textbf{Department of Physics, Yale University, New Haven, Connecticut 06520, USA}}}\newline
\noindent\textit{\small{\textit{$^{e}$}\textbf{Department of Chemical Engineering, Columbia University, New York, New York 10027, USA}}}\newline

\noindent\textit{\small{\textbf{Received Xth XXXXXXXXXX 20XX, Accepted Xth XXXXXXXXX 20XX\newline
First published on the web Xth XXXXXXXXXX 200X}}}

\noindent \textbf{\small{DOI: 10.1039/b000000x}}
\vspace{0.6cm}

\noindent \normalsize{Key features of the mechanical response of 
amorphous particulate materials, such as foams,
emulsions, and granular media, to applied stress are determined by the 
frequency and size of particle rearrangements that occur as the 
system transitions from one
mechanically stable state to another. This work describes
coordinated experimental and computational studies of bubble rafts,
which are quasi-two dimensional systems of bubbles confined to the
air-water interface. We focus on small mechanically stable clusters
of four, five, six, and seven bubbles with two different sizes with 
diameter ratio $\sigma_L/\sigma_S \simeq 1.4$.
Focusing on
small bubble clusters, which can be viewed as subsystems of a larger system, 
allows us to investigate the full ensemble of clusters
that form, measure the respective frequencies with which the clusters occur,
and determine the form of the bubble-bubble interactions.  We emphasize
several important results. First, for clusters with $N > 5$
bubbles, we find using discrete element simulations that short-range 
attractive interactions between bubbles give rise
to a larger ensemble of distinct mechanically stable clusters compared to 
that generated by long-range attractive interactions.
The additional clusters in systems with short-range attractions possess 
larger gaps between pairs of neighboring bubbles on the periphery 
of the clusters.  The ensemble of bubble clusters observed in experiments 
is similar to the ensemble of clusters with long-range 
attractive interactions. We also compare the frequency with which 
each cluster occurs in simulations and experiments. We find that the 
cluster frequencies are extremely sensitive to the protocol used 
to generate them and only weakly correlated to the energy of 
the clusters.}
\vspace{0.5cm}
 \end{@twocolumnfalse}
  ]

\section{Introduction}
\label{sec:Introduction}

One of the most important, open questions in condensed matter physics
involves understanding the mechanical response of amorphous materials,
which do not possess long-range positional or bond-orientational
order.  In crystalline materials, the mechanical response is
well-described by the motion and interaction of topological defects,
such as dislocations. In contrast, it is much more difficult to
identify the relevant defects that control the mechanical response in
amorphous materials~\cite{sandfeld}.  Disordered materials span a wide
range of length scales from atomic systems, such as metallic
glasses~\cite{greer2009metallic} and ceramics~\cite{kingery1960}, to
colloidal suspensions~\cite{SWS07,CSSBW10}, and macroscopic
particulate materials, such as foams~\cite{DTM01,D04,LCD04,LKXO08},
emulsions~\cite{pine}, and granular
matter~\cite{UB08,KA13,JWS14,LRLD15}.

There are several common features that characterize the mechanical
response of disordered materials. For small strains, or applied
stresses, these materials have an initial elastic-like response during
which the stress is proportional to strain, particle-scale
rearrangements are rare, and the system is essentially reversible upon
increasing and decreasing the strain. Above a characteristic applied
stress or strain, which is typically much smaller than that for
crystalline materials, particle rearrangements begin to occur
frequently, and the material effectively flows with large fluctuations
in stress (or energy).  Stress fluctuations are often characterized by
periods of increasing stress followed by sudden decreases, referred to
as stress drops~\cite{LTD02,PD03,PALB05,DW15,meng} or
avalanches~\cite{JSSA99}. During flow, particle rearrangements can
localize, which gives rise to shear
banding~\cite{KD03,WKD06a,KSD07,GCOA08,DGM13}. 

Many theoretical descriptions of disordered solids have employed
models that are relatively insensitive to the details of the particle
interactions and have focused on the concept of flow defects to
understand deformation of disordered materials during applied
loading~\cite{S77, A79}. For example, shear transformation zone (STZ)
theories describe the creation and annihilation of regions in a
material that experience large non-affine particle
displacements~\cite{FL98}.  STZ theories have been applied to a wide
range of materials, including amorphous metals, colloidal glasses, and
earthquake faults
\cite{FL98,ELP_PRL_2003,LP03,L04,P05,M06,L06,BLP07a,BLP07b,L08}. Similar
descriptions include weak-zone and soft-spot theories, which attempt
to identify regions in a material where non-affine deformation will
occur due to mechanical instabilities~\cite{ML11,SLRR14,tong}.

Despite the successes of such theoretical descriptions, the influence
of particle interactions on mechanical response can be significant, for
example, when considering the structure and dynamics of foams. In
two-dimensional (2D) dry foams, bubbles are polygonal and form a
connected network of bubble edges.  In this case, one can precisely
define topological rearrangements (T1 events) associated with the
switching of bubble neighbors, where one of the polygonal edges
decreases to zero and then a new edge is created. T1 events can be
isolated and localized, or occur as multiple, correlated
rearrangements~\cite{PF09,KK12}. In fact, one event at a given
location in the system at a given time can trigger another at a
different location later in time.  Theoretical descriptions of the
mechanical properties of dry foams are typically formulated in terms
of the activation of T1 and other topological
events~\cite{TSDK99,JSSA99,LTD02,KD03,D04,PALB05,WKD06a,DS06,KSD07,LKXO08,KA13,DW15}.

In the wet-foam limit, bubbles do not form fully connected polyhedral
networks and thus wet foams behave more like particulate media, which
do not form confluent networks of interparticle contacts. However,
bubble systems do possess attractive interactions, and
thus each bubble is in contact with at least some of its neighbors~\cite{VF01}.
Particulate systems with purely repulsive interactions, such as dry
granular media, possess a finite fraction of ``rattler'' particles,
which are disconnected from the force-bearing network of
contacts. Rattler particles complicate the mechanical response of
disordered solids because they can intermittently join and exit the
force-bearing backbone.

In this article, we describe a series of coordinated experimental and
computational studies of the mechanically stable clusters that occur
in bubble rafts, which are layers of mm-sized bubbles that float on
the surface of water and mimic wet foams.  These systems are mainly
two-dimensional, which allows us to accurately determine the positions
of bubble centers and contacts between bubbles. Another advantage of
bubble rafts is that they are athermal ({\it i.e.} the forces
generated from thermal fluctuations are small compared to surface
tension), so that we can determine the mechanically stable clusters of
small bubble clusters in the absence of thermal fluctuations. In
addition, in contrast to other athermal systems such as granular
media, bubble rafts do not possess static frictional interactions, for
which enumeration of mechanically stable packings is much more
difficult~\cite{shen}.

Attractive forces between bubbles include long- and short-range
contributions. First, the bubbles deform the surface of the water,
which creates a long-range capillary-like
interaction~\cite{KN00,KD01}. However, the surface deformation and
resulting form of the attractions depend on the number, size, and
relative positions of the bubbles in a cluster. In addition to this
long-range attractive force, the bubbles experience short-range
interactions when bubbles come into contact arising from surface
tension.  Note that this interaction depends on the particular
trajectories of interacting bubbles. For example, the interactions are
different when bubbles are approaching each other compared to when
they are moving away from each other.

\begin{figure*}
	\begin{center}
		\includegraphics[width=0.7\textwidth]{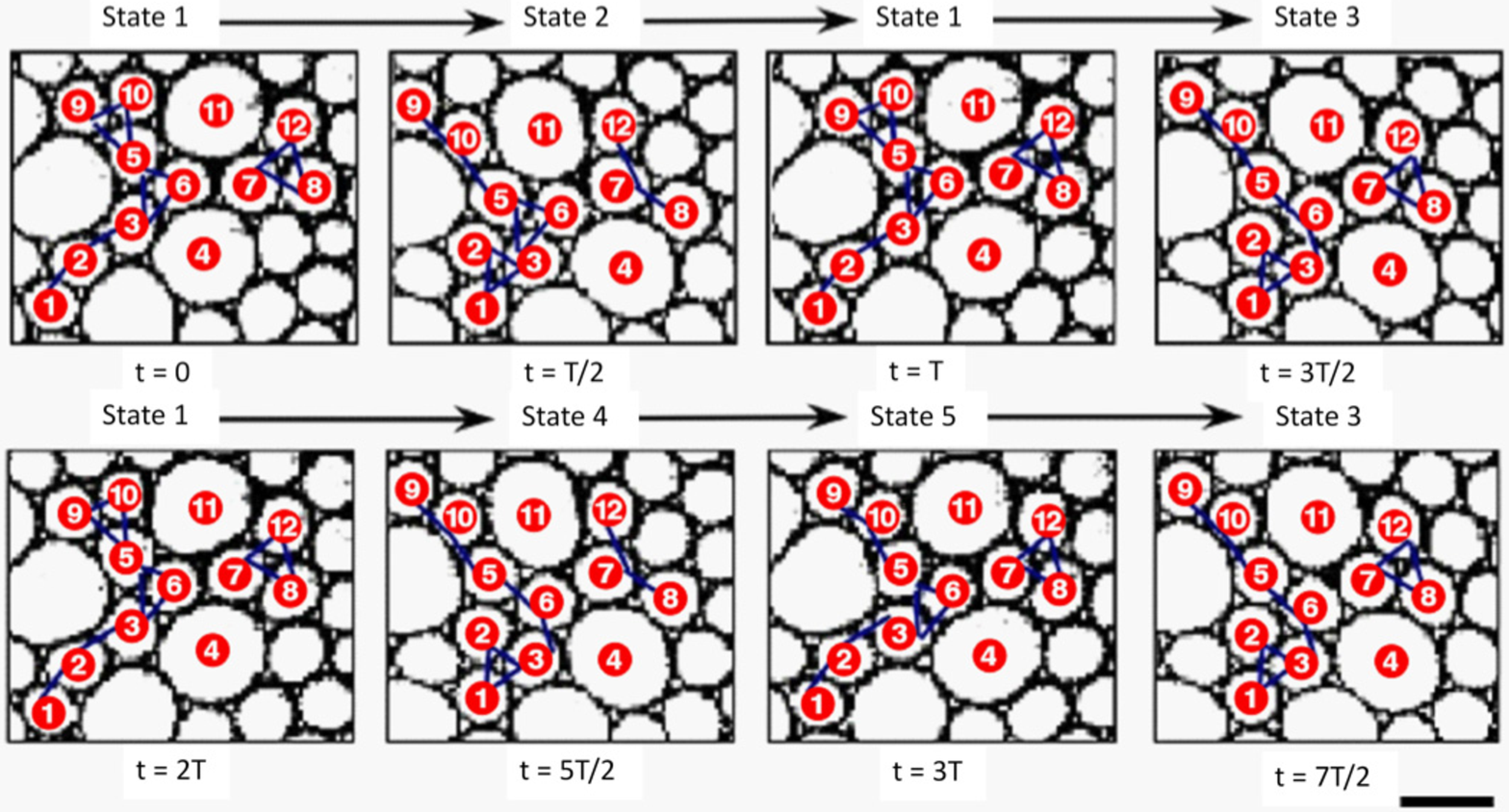}
	\end{center}
\caption{A series of images taken from the central region of a bubble
raft undergoing oscillatory shear in previous experiments~\cite{LKXO08}. 
The bubble diameters are in the
range $2.5$ to $5.3~{\rm mm}$. (A $5\ {\rm mm}$ scale bar is given in the 
lower right corner of the figure.)  The
driving amplitude and frequency were $28$~mm and $0.2$~Hz (period $T
= 5\ {\rm s}$), respectively, which corresponds to a root-mean-square (rms) 
strain and
strain rate of $0.22$ and $0.04\ {\rm s^{-1}}$, respectively. The
images are taken at the turning points of the oscillatory driving
(2.5 s apart).  The same 12 bubbles are labeled in each image. The
bubble motions suggest defining four clusters of three bubbles, (5,
9, 10), (1, 2, 3), (3, 5, 6), and (7, 8, 12), where the three-bubble
clusters can exist in linear or triangular arrangements (blue
lines).}
	\label{fig:rearrange}
\end{figure*}

In previous work on bubble rafts~\cite{LKXO08}, we characterized
bubble rearrangements induced during cyclic shear in large systems in
terms of the relative motions of bubbles in three-bubble clusters, as
well as T1 events that involve four-bubble clusters. We showed that in
bubble rafts, there are a number of bubble rearrangements that cannot
be decomposed into a set of T1 events exclusively because they involve
clusters of three bubbles with triangular or linear topologies. In
Fig.~\ref{fig:rearrange}, we show eight images at the turning points
during applied oscillatory strain over $3.5$ cycles. The bubble
configurations suggest defining four clusters of three small bubbles
each.  We label the bubbles in the four clusters with indexes, $(5, 9,
10)$, $(1, 2, 3)$, $(3, 5, 6)$, and $(7, 8, 12)$. During the cyclic
shear, the three-bubble clusters transition between linear and
triangular structures. If one considers more complex structures
involving bubbles of different sizes, the linear and triangular
three-bubble clusters form the basic building blocks for clusters of
four, five, six, and larger bubble clusters.

Motivated by these preliminary studies, here we focus on small bubble
clusters ($N=4$, $5$, $6$, and $7$) with bubbles of two different sizes
(identified as large (L) and small (S) with diameter ratio
$\sigma_L/\sigma_S \simeq 1.4$). For a given $N=N_S+N_L$, we vary the
number of small and large bubbles, $N_S$ and $N_L$, respectively, in
each cluster in the range $N_S,N_L\le 3$. For $N=7$, we studied clusters with 
$N_L=4$ and $N_S=3$. Focusing on small bubble
clusters allows us to enumerate all mechanically stable clusters and
calculate the frequency with which they occur in both experiments and
simulations.  We will compare three important quantities in
experiments and simulations: (a) the number of distinct mechanically
stable clusters; (b) bubble separations in each distinct cluster; and (c)
the frequency with which each cluster occurs.  For four- and
five-bubble clusters, we observe that these three quantities are
largely independent of the form of the bubble-bubble
interactions. However, we find that the properties of six and larger
bubble clusters depend on the range and strength of the bubble
attractive interactions. Thus, the comparison of properties of mechanically
stable six-bubble clusters in experiments and simulations allows us to
calibrate the bubble-bubble interaction potential in bubble rafts.

The remainder of this article is organized as follows. In
Sec.~\ref{experimental-details}, we describe the experimental setup
and methods employed to generate the bubble clusters. In
Sec.~\ref{sim-details}, we describe the bubble-bubble interaction
potential and protocol for generating the bubble clusters in
simulations. In Secs.~\ref{sec:results} and ~\ref{disc}, we summarize
the results from the simulations and experiments and discuss future
research directions.

\section{Experimental System: Bubble Rafts}
\label{experimental-details}

Bubble rafts consist of a single layer of bubbles at the air-water
interface.  They represent the limit of wet foams and have been used
to model defects in atomic solids~\cite{B42,BN47}. Bubble rafts offer
a number of advantages as an experimental system for studying bubble
clusters. First, bubbles are formed individually by flowing compressed
nitrogen gas through a thin needle into a bubble solution. The
combination of needle size, flow pressure, and bubble depth gives
precise control over the bubble size. For our work, we used a standard
bubble solution consisting of $80\%$ DI water, $15\%$ glycerin, and
$5\%$ bubble solution by volume. The bubble solution is Alkaline Liquid Detergent from Contrex. This mixture
generates bubble rafts that are stable for $1$-$2$ hours. We 
work with bubbles having diameters in the range from $0.5\ {\rm mm}$
to a few mm. The ability to control the bubble size allows for the
creation of monodisperse and bidisperse size distributions. The
quasi-two dimensional nature of the system enables accurate tracking
of all bubbles in the system.

\begin{figure}
	\begin{center}
		\includegraphics[width=0.9\columnwidth]{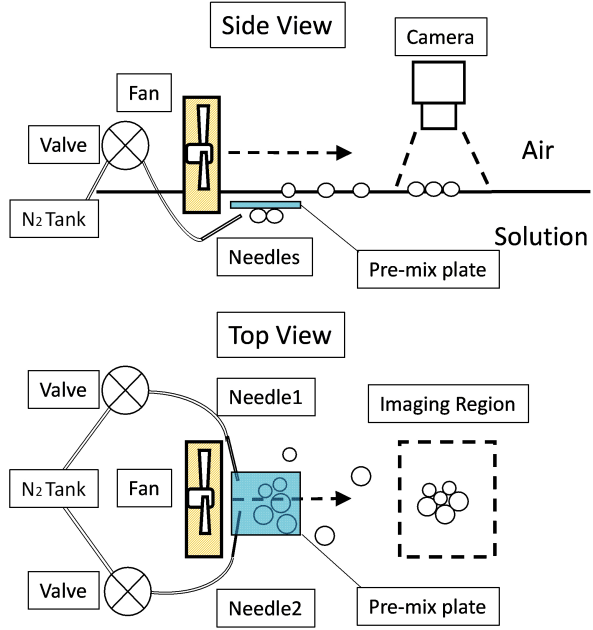}
	\end{center}
\caption{Schematics of the (top panel) side and (bottom panel) top views of the
experimental setup for generating large numbers of small clusters of
bubbles. Bubbles with two different sizes are created using needles
with different inner diameters. A computer-controlled fan is used to
move the bubbles from the pre-mix plate to the imaging region. The clusters form as
the bubbles move away from the fan and decrease in speed.}
\label{fig:bubgen}
\end{figure}

The experimental setup used to generate random clusters of bubbles is
illustrated in Fig.~\ref{fig:bubgen}. Our work has focused on clusters
formed from bubbles with two sizes with diameter ratio
$\sigma_L/\sigma_S = 1.45 \pm 0.03$.  We employ two injection needles
with different inner diameters to generate the two bubble sizes. The
injection speed is controlled by the mass flow rate from the
compressed nitrogen gas tank. By placing a cover slide between the
needles and at the air-water interface, we create a mixing region
that introduces randomness into the cluster formation process. Additional
fluctuations are included by exposing the system to an air current
using a computer-controlled fan with a time dependent on-off
oscillation. Images of the clusters are captured down stream of the
generation region and analyzed to determine the positions of all
bubbles in the clusters.

We identified several key features of the experimental bubble cluster
formation process. First, bubble clusters containing $N=1$, $2$, or
$3$ bubbles commonly form directly outside the pre-mix plate region.
These clusters immediately combine and rearrange to form new clusters
with size $N > 2$.  It is much rarer for initial clusters with $N > 4$
to form directly near the pre-mix plate region. As we will show below,
the protocol for bubble cluster generation can strongly affect the
frequencies with which distinct mechanically stable clusters form. To vary the 
protocol, we considered two pre-mix plate regions with different areas: protocol $A$
with area $1.9\ {\rm cm} \times 1.4\ {\rm cm}$ and protocol
$B$ with area $1.2\ {\rm cm} \times 1.4\ {\rm cm}$. Unless noted
otherwise, the reported results were obtained using protocol A. The
different mixing geometries created an observable change in the frequencies
of precursor clusters of size $N < 6$, which results in changes to the
frequencies of larger bubble clusters.

\section{Simulation Details}
\label{sim-details}

For bubble rafts undergoing confined shear flows, the assumption has
been that only repulsive contact forces and viscous forces are
necessary to describe their macroscopic response
~\cite{TSDK99,durian}. However, bubble-bubble adhesion and capillary
forces~\cite{KN00,KD01} can give rise to strong attractive
interactions between bubbles. The role of attractive interactions in
determining transitions between mechanically stable clusters,
especially in the wet foam limit, has not been investigated in detail.

Here, we will model interactions between bubbles in bubble rafts 
using the generalized Lennard-Jones (LJ) pairwise potential:
\begin{equation}
\label{eq:uljmn}
V_{\rm LJ}^{\rm mn}(r_{ij}) = \epsilon_{ij} \left[ 2^{\frac{m}{6}} \frac{n}{m-n} \left(\frac{\sigma_{ij}}{r_{ij}}\right)^m - 2^{\frac{n}{6}} \frac{m}{m-n} \left(\frac{\sigma_{ij}}{r_{ij}}\right)^n \right].
\end{equation}
When $m=12$ and $n=6$ (denoted as $(12,6)$), Eq.~\ref{eq:uljmn} reduces
to the standard Lennard-Jones interaction potential:
\begin{equation}
\label{eq:ulj}
V_{\rm LJ}(r_{ij}) = 4\epsilon_{ij} \left[ \left(\frac{\sigma_{ij}}{r_{ij}}\right)^{12} -  \left(\frac{\sigma_{ij}}{r_{ij}}\right)^6 \right],
\end{equation}
where $r_{ij}$ is the separation between the centers of bubbles $i$
and $j$, $\sigma_{ij} = (\sigma_i+\sigma_j)/2$, $\sigma_i$ is the
diameter of bubble $i$, the minimum in the pair potential $V_m =
-\epsilon_{ij}$ occurs at $r_m = 2^{1/6} \sigma_{ij}$, and the energy
parameters satisfy the mixing rule, $\epsilon_{ij} \sim
\sigma_i\sigma_j /(\sigma_i+\sigma_j)$, which follows the
Johnson-Kendall-Roberts (JKR) theory for contact mechanics
~\cite{johnson:1971,israelachvili:2011}.  The total potential energy
per bubble is given by $U/N = N^{-1} \sum_{ij} V_{\rm LJ}^{\rm
  mn}(r_{ij})$. Energies, lengths, timescales, and temperatures are
given in units of $\epsilon_{LL}$, $\sigma_{LL}$,
$\sigma_{LL}\sqrt{m/\epsilon_{LL}}$, and $\epsilon_{LL}/k_B$,
respectively, where the Boltzmann constant $k_B$ and bubble mass $m$
are set to unity.

\begin{figure}
\includegraphics[width=0.8\columnwidth]{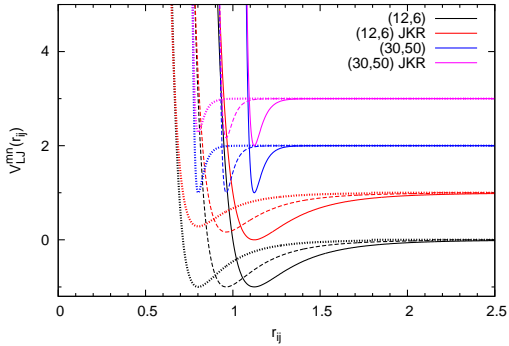}
\caption{Illustration of the generalized Lennard-Jones pairwise 
potential (Eq.~\ref{eq:uljmn}) used in the 
simulations to model the bubble-bubble interactions in bubble rafts.  
We consider 
long-range $(m,n)=(12,6)$ and short-range $(30,50)$ attractive 
LJ interactions with $\epsilon_{ij} =\epsilon_{LL}$ and JKR mixing rules. For
each potential, the solid, dotted, and dashed lines indicate the 
interactions between two
large bubbles, two small bubbles, and a small and
large bubble, respectively. The potentials have been shifted vertically for
clarity.} 
\label{fig:poteg}
\end{figure}

For the generalized LJ potential in Eq.~\ref{eq:uljmn}, we fix the
location of the minimum at $r_m$, but change the exponents $m$ and $n$
to tune the width of the attractive well,
$\Delta(m,n)=2^{1/6}\sigma_{ij}[
  ((m+1)/(n+1))^{1/(m-n)}-(n/m)^{1/(m-n)}]$, which is defined as the
distance between the zero of the potential ($V_{\rm LJ}^{\rm
  mn}(r_{ij})=0$) and the inflection point of the attractive tail
($d^2 V_{\rm LJ}^{\rm mn}(r_{ij})/dr_{ij}^2=0$). For the Lennard-Jones
potential, the width $\Delta(12,6)=0.24\sigma_{ij}$, while it
decreases to $\Delta(30,50)=0.06\sigma_{ij}$ for larger values of $m$
and $n$. This allows us to tune the interactions from long-range
$(12,6)$ to short-range attractions $(30,50)$. We can also
independently vary the strength of the attractions by tuning
$\epsilon_{ij}$. For these studies, we used $\epsilon_{ij} =
\epsilon_{LL}$ and the JKR mixing rule $\epsilon_{ij} \sim
\sigma_i\sigma_j /(\sigma_i+\sigma_j)$, as well as stronger attractions
with $\epsilon_{ij} \sim \sigma_{ij}^3$.  For calculations of the 
inter-bubble forces and potential energy, we cut off the interactions beyond
$r_c=2.5\sigma_{ij}$~\cite{dyre}. We illustrate the long-range
$(12,6)$ and short-range $(30,50)$ attractive potentials with
$\epsilon_{ij} = \epsilon_{LL}$ and the JKR mixing rule in
Fig.~\ref{fig:poteg}.

We use two protocols to generate bubble clusters in the simulations:
(1) random initialization of all bubbles at once and (2) sequential
addition of bubbles one by one in the simulation cell. In protocol
$1$, we initialize $N$ bubbles with random positions in a small
simulation cell with velocities randomly chosen from a Gaussian
distribution at temperature $T$.  We then decrease the temperature to
zero linearly in time at different cooling rates $R = dT/dt$.  As $T
\rightarrow 0$, the bubbles (interacting via Eq.~\ref{eq:uljmn})
aggregate into clusters.  We employed protocol $1$ with both open boundaries 
and closed boundaries with repulsive walls. The only difference between 
the two boundary conditions is that bubbles in the system with the closed boundaries were 
more likely to form a single cluster. 

In protocol (2), we insert {\em one} bubble at a time with kinetic
energy $k_B T$ at a random position into the simulation cell with
closed boundaries. We then cool the system to zero temperature using a
linear ramp with rate $R$. The next bubble is inserted randomly into
the simulation cell (without overlapping the previously placed, fixed
bubble) and then the system is cooled to $T=0$ at rate $R$. This
process is continued until all $N$ bubbles have been inserted. We can
vary the insertion sequence over all possible combinations of large
and small bubbles at fixed $N_L$ and $N_S$ for a given $N$. For
example, for $N=4$, $N_L=1$, and $N_S=3$, we consider $LSSS$, $SLSS$,
$SSLS$, and $SSSL$ insertion orders for the small and large bubbles.

For each protocol and $N$, we performed a large number $M=10^3$ (for
closed boundaries) or $10^4$ (for open boundaries) runs with different
initial conditions. We also varied the cooling rate over two orders of
magnitude: $10^{-3} \le R \le 10^{-1}$. After applying the linear
cooling ramp to $T=0$, conjugate gradient energy minimization was
employed to ensure that the final $T=0$ clusters are mechanically
stable. We distinguish the mechanically stable clusters using the
networks of interparticle contacts and determine the frequency with
which each distinct cluster occurs.

\section{Results}
\label{sec:results}

We present the results in two sections. First, we focus on the
structure ({\it i.e.} the arrangement and spacing between bubbles) of
the mechanically stable clusters. To compare the distinct bubble
clusters obtained in experiments and simulations, we define the
distance in configuration space as $\Delta R = \sqrt{\sum_{i=1}^N
  [(x_i^e- x_i^s)^2 + (y_i^e-y_i^s)^2]}$, where
$(x^{e,s}_i,y^{e,s}_i)$ gives the geometric center of bubble $i$ for a
given cluster from experiments and simulations, respectively.  Second,
we consider the frequency with which each distinct cluster occurs. We
compare the cluster frequencies from experiments and simulations for
different interaction potentials and cluster-generation protocols.

\subsection{Structure of mechanically stable clusters}
\label{structure}

\begin{figure}
\begin{center}
\includegraphics[width=0.9\columnwidth]{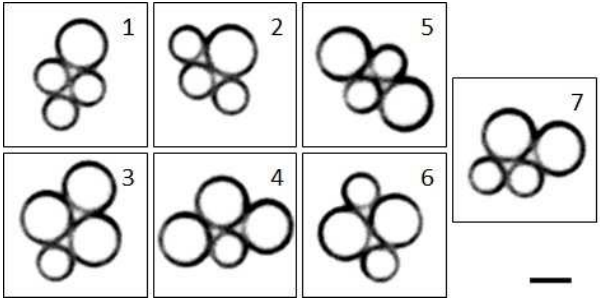}
\end{center}
\caption{The $7$ distinct mechanically stable clusters 
with $N=4$ bubbles obtained from experiments and simulations. 
There are two clusters with $N_S=3$ and $N_L=1$, 
three clusters with $N_S=2$ and $N_L=2$, and two clusters 
with $N_S=1$ and $N_L=3$.  All clusters possess an isostatic 
number of bubble contacts, $N_c=N_c^{\rm iso}=5$.
The clusters are indexed (upper right corner of each panel) 
the same in simulations and experiments. The $2\ {\rm mm}$ scale bar for 
each panel is in the lower right corner of the figure.} 
\label{fig:exp4}
\end{figure}

\begin{figure}
\begin{center}
\includegraphics[width=0.9\columnwidth]{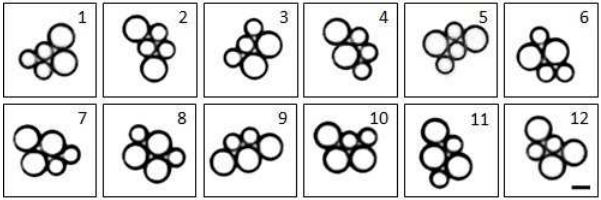}
\end{center}
\caption{The $12$ distinct mechanically stable clusters of $N=5$ bubbles 
with $N_L=2$ and $N_S=3$ and $N_L=3$ and $N_S=2$ found in experiments
and simulations. All clusters are isostatic with $N_c=N_c^{\rm iso}=7$ 
bubble contacts. The cluster index is given in upper right corner of 
each panel. The $2\ {\rm mm}$ scale bar for each panel is 
given in the lower right corner of the figure.} 
\label{fig:exp5}
\end{figure}

We first describe the results for four- and five-bubble clusters.  For
these systems, we focus on combinations of large and small bubbles
that can serve as precursor clusters for $N = 6$ with three small and
three large bubbles. In Figs.~\ref{fig:exp4} and~\ref{fig:exp5}, we
show the $7$ distinct mechanically stable clusters for $N=4$ (with
$N_S=3$, $N_L=1$; $N_S=2$, $N_L=2$; and $N_S=1$, $N_L=3$) and $12$
distinct mechanically stable clusters for $N=5$ (with $N_S=3$, $N_L=2$
and $N_S=2$, $N_L=3$) that are observed in both experiments and
simulations. All of the $N=4$ and $5$ bubble clusters are isostatic
with $N_c=N_c^{\rm iso}=2N-3$ bubble contacts~\cite{witten}. Note that
mechanically stable clusters must possess $N_c \ge N_c^{\rm iso}$. For
all forms of Eq.~\ref{eq:uljmn} that we studied, the simulations
generate the same $7$ mechanically stable clusters for $N=4$ and same
$12$ mechanically stable clusters for $N=5$.  The $N=4$ and $5$
clusters are compact such that all nearest neighbor bubbles are in
contact with each other.  The distance in configuration space $\Delta
R$ between a given cluster obtained in experiments and that obtained
in simulations is small, with $\Delta R/\sigma_{LL} < 0.15$.

\begin{figure}
\begin{center}
\includegraphics[width=0.9\columnwidth]{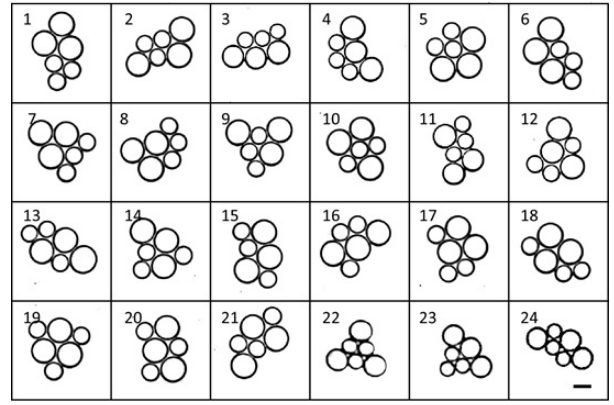}
\end{center}
\caption{The $24$ mechanically stable $N=6$ clusters (with $N_L=3$ 
and $N_S=3$) obtained from the bubble raft experiments. The unique cluster 
index is given in the upper left corner. The $2\ {\rm mm}$ scale bar for 
each panel is given in the lower right corner of the figure.} 
\label{fig:exp6}
\end{figure}

\begin{figure}
\begin{center}
\includegraphics[width=1\columnwidth]{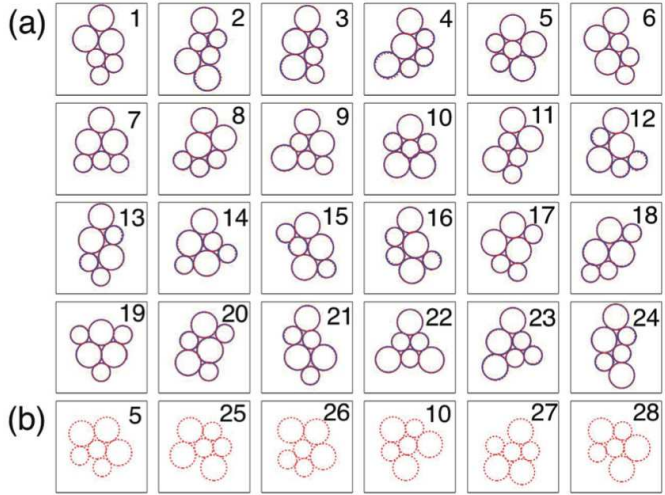}
\end{center}
\caption{(a) $24$ of the distinct mechanically stable clusters of
$N=6$ bubbles (with $N_L=3$ and $N_S=3$) obtained from simulations
using the $(12,6)$ generalized LJ potential with JKR mixing rules,
protocol $1$, and $R=10^{-1}$.  Clusters $1$-$24$ (except $5$ and $10$) 
are isostatic with $N_c=9$ bubble contacts; clusters $5$ and $10$ have 
$10$ bubble contacts.  The cluster index is given by the
number in the upper right corner. (b) When using short-ranged
attractive interactions, {\it e.g.} $(30,50)$, in the simulations,
clusters $5$ and $10$ change and four additional clusters labeled
$25$-$28$ are generated. In each panel, the bubbles are outlined as
solid and dotted lines for the clusters obtained by experiments
and simulations, respectively.}
\label{fig:n6sim}
\end{figure}

For $N=6$, we observe $24$ distinct mechanically stable clusters (with
$N_L=3$ and $N_S=3$) in the bubble raft experiments as shown in
Fig.~\ref{fig:exp6}. In simulations (using protocol $1$), we also
observe $24$ mechanically stable clusters that are similar to those
observed in the experiments when we consider the long-range $(12,6)$
LJ potential with the JKR mixing rules (Fig.~\ref{fig:n6sim} (a)). In
this case, $22$ of the clusters ($1$-$24$ except $5$ and $10$) are
isostatic with $9$ bubble contacts, however, clusters $5$ and $10$ are
hyperstatic with $10$ bubble contacts. When the generalized LJ
potential is tuned from long- $(12,6)$ to short-ranged $(30,50)$
attractive interactions, the structures of clusters $5$ and $10$
change significantly and four additional clusters ($25$-$28$) are
generated (Fig.~\ref{fig:n6sim} (b)). The difference between cluster
$5$ generated using the $(12,6)$ potential and cluster $5$ generated
from the $(30,50)$ potential is that a gap opens between two small
bubbles on the periphery of the cluster that were in contact for the
$(12,6)$ potential. Similarly, two large bubbles that were in contact
in cluster $10$ for the $(12,6)$ potential move apart in cluster $10$
for the $(30,50)$ potential.  Clusters $25$ and $26$ ($27$ and $28$)
that occur for the $(30,50)$ potential are similar in structure to
cluster $5$ ($10$), with gaps between different pairs of bubbles on
the periphery.

\begin{figure}
	\includegraphics[width=0.9\columnwidth]{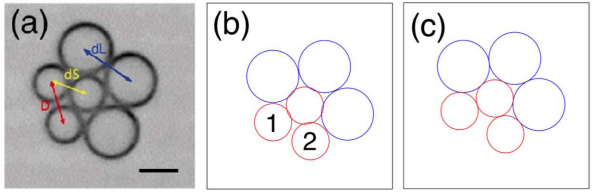}
\caption{Mechanically stable bubble cluster $5$ for $N=6$
(with $N_L=3$ and $N_S=3$) obtained from (a) experiments and
simulations with the (b) long-ranged $(12,6)$ ($\epsilon_{ij}
\sim \sigma_{ij}^3$) and (c) short-ranged $(30,50)$ generalized
Lennard-Jones potential. $D$ gives the separation between small
bubbles $1$ and $2$ on the periphery of cluster $5$. $dS$ and $dL$
give the separations between two contacting small and large bubbles,
respectively. The scale bar in (a) is $2\ {\rm mm}$.}
\label{fig:gapimage}
\end{figure}

\begin{figure}
	\includegraphics[width=0.9\columnwidth]{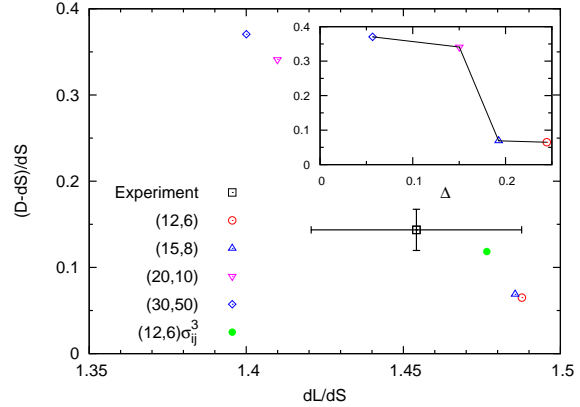}
\caption{Gap distance $(D - dS)/dS$ between small bubbles $1$ and $2$
in cluster $5$ (Fig.~\ref{fig:gapimage}) as a function of the ratio
$dL/dS$ from experiments and simulations (using protocol $1$) with
different $(m,n)$ and the JKR mixing rule. We also show the gap
distance obtained from the $(12,6)$ potential with $\epsilon_{ij}
\sim \sigma_{ij}^3$.  The inset shows $(D-dS)/dS$ versus the width
of the interaction potential $\Delta(m,n)$ from the simulations that
are shown in the main figure.}
\label{fig:gap}
\end{figure}

In Fig.~\ref{fig:gapimage}, we compare images of bubble cluster $5$ obtained in
experiments and simulations with long- and short-range attractions.
We characterize the inter-bubble separations in Fig.~\ref{fig:gap}, by
plotting the normalized gap distance $\delta = (D - dS)/dS$ between
bubbles $1$ and $2$ versus $dL/dS$ for cluster $5$, where $D$ is the
center-to-center distance between bubbles $1$ and $2$ and $dS$ ($dL$)
is the center-to-center distance between two contacting small (large)
bubbles.  In the experiments, we find that the normalized gap distance
is $\delta = 0.14 \pm 0.02$.  In simulations using protocol $1$ and
JKR mixing rules, the normalized gap distance is different for short-
versus long-ranged attractive potentials. We find a relatively large
normalized gap distance $\delta \sim 0.35$ for short-range attractions
($(30,50)$ and $(20,10)$), whereas $\delta \sim 0.05$ for long-range
attractions ($(15,8)$ and $(12,6)$). As shown in the inset to
Fig.~\ref{fig:gap}, the normalized gap distance increases abruptly as
the range of the potential is decreased below $\Delta \approx 0.2$.

The normalized gap $\delta$ between bubbles $1$ and $2$ in cluster $5$
from experiments falls between the two extremes from simulations 
using short- and long-ranged attractions shown in the inset to
Fig.~\ref{fig:gap}, though it closer to the long-range values. Therefore,
we also considered the impact of the interaction strength on the
normalized gap. We find that increasing the interaction strength for
the long-range $(12,6)$ potential moves the gap distance between
bubbles $1$ and $2$ in cluster $5$ closer to the experimental
value. (See the result for $(12,6)$ $\epsilon_{ij} \sim \sigma_{ij}^3$ in
Fig.~\ref{fig:gap}.)  In Fig.~\ref{fig:gapimage}, we compare the images
of cluster
$5$ obtained from (a) experiments with cluster $5$ obtained from
simulations with (b) long-ranged attractions ($(12,6)$ and
$\epsilon_{ij} \sim \sigma_{ij}^3$) and (c) short-ranged attractions
($(30,50)$ and JKR mixing rules).

\begin{figure}
	\includegraphics[width=0.9\columnwidth]{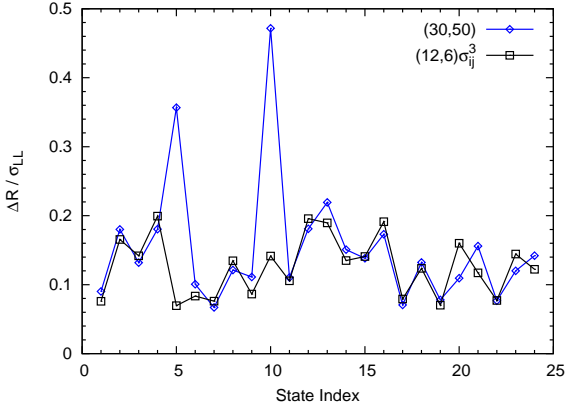}
\caption{The distance in configuration space $\Delta R/\sigma_{LL}$
between a distinct cluster obtained in simulations and that obtained
in experiments for $N=6$. For the simulations, we used the $(12,6)$
long-ranged potential with $\epsilon_{ij} \sim \sigma_{ij}^3$ mixing
rules and the short-ranged $(30,50)$ potential.  The $(12,6)$
potential with JKR mixing rules yields similar results. One can see
that clusters $5$ and $10$ are the most sensitive to the form of the
potential.}
\label{fig:dr}
\end{figure}

To quantify the impact of the range and strength of the attractive
bubble interactions on the structure of all bubble clusters, we
computed the distance in configuration space $\Delta R/\sigma_{LL}$
between the $N=6$ clusters obtained from simulations using the
$(12,6)$ and $(30,50)$ generalized Lennard-Jones potential and the
clusters obtained in experiments (Fig.~\ref{fig:exp6}). We find that
the distance in configuration space between each cluster obtained in
simulations with long-range attractions $(12,6)$ and the corresponding
cluster obtained in experiments satisfies $\Delta R/\sigma_{LL} <
0.2$, as shown in Fig.~\ref{fig:dr}.  In contrast, $\Delta
R/\sigma_{LL}$ can be much larger when comparing the clusters from
experiments those obtained from simulations using the short-ranged
$(30,50)$ potential, {\it e.g.} $\Delta R/\sigma_{LL} \sim 0.5$ for
cluster $10$.

\begin{figure}
	\begin{center}
	\includegraphics[width=0.9\columnwidth]{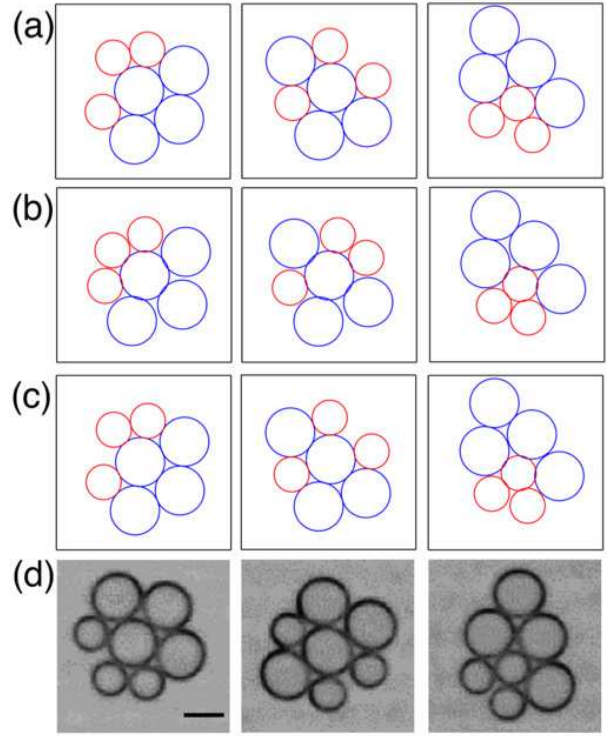}
	\end{center}
\caption{Three example classes of bubble clusters that illustrate the
impact of the interaction potential on gaps between bubbles on the
periphery of clusters. (a) Three clusters generated using the
$(30,50)$ generalized Lennard-Jones potential with short-ranged
attractions and JKR mixing rules. Each cluster has a pair of bubbles
on the periphery with a large gap. (b) Clusters generated using the
$(12,6)$ potential with long-range attractions and JKR mixing
rules. None of the pairs of bubbles on the periphery possess a
gap. (c) Clusters generated using the $(12,6)$ potential with
$\epsilon_{ij} \sim \sigma_{ij}^3$. Some of the clusters have bubble
pairs on the periphery with gaps, while others do not. (d) Bubble clusters
from experiments exhibit behavior similar to the clusters
in (c) generated using the $(12,6)$ potential with $\epsilon_{ij}
\sim \sigma_{ij}^3$. The scale bar in panel (d) is $2\ {\rm mm}$.}
\label{fig:exp7}
\end{figure}

To further investigate the role of the bubble interaction potential on
the types of mechanically stable bubble clusters that occur, we performed
preliminary studies of $N = 7$ clusters with $N_L = 4$ and $N_S = 3$.
We find that there are more than hundred distinct mechanically stable
clusters for $N=7$ with $N_L=4$ and $N_S=3$. For $N=6$, we
found that the simulations using the $(12,6)$ and $(30,50)$ potentials
generated $N_c = 22$ mechanically stable clusters that were the
same. Two clusters had similar
topologies for the $(12,6)$ and $(30,50)$ potentials, but those for
the $(30,50)$ potential had more gaps between bubbles on the periphery 
of the clusters. In addition, the
$(30,50)$ potential generated four clusters that were not found
for the $(12,6)$ potential. (See Fig.~\ref{fig:n6sim}.)  For $N=7$, we
find increasing disparities in structure between the ensemble of
mechanically stable clusters generated using the $(12,6)$ and
$(30,50)$ potentials.  For example, we find that the fraction of
distinct bubble clusters that are the same for the $(12,6)$
and $(30,50)$ potentials decreases for $N=7$.  In Fig.~\ref{fig:exp7}, we show
representative clusters for $N=7$ from simulations using the $(30,50)$
potential and the $(12,6)$ potential with the JKR and $\epsilon_{ij} \sim
\sigma_{ij}^3$ mixing rules.  The clusters obtained from the
short-ranged $(30,50)$ potential typically possess large gaps between
small bubbles on the periphery (panel (a)).  In contrast, clusters
obtained from the long-ranged $(12,6)$ potential do not typically
possess large gaps between bubbles on the periphery (panel (b)).  For
the $(12,6)$ potential with $\epsilon_{ij} \sim \sigma_{ij}^3$, there
is a mixture of clusters with and without large gaps (panel (c)),
which mimics the structure of the clusters found in experiments (panel
(d)).

\subsection{Frequency of mechanically stable clusters}
\label{frequency}

In the previous section, we showed that the structure of the distinct
mechanically stable bubble clusters in experiments was similar to the
structure of the clusters generated from simulations using long-range
$(12,6)$ attractive interactions. In this section, we will investigate
to what extent the frequencies with which each distinct bubble cluster
occurs is sensitive to the bubble interaction potential and the
protocol used to generate the bubble clusters.

\begin{figure}
	\includegraphics[width=0.9\columnwidth]{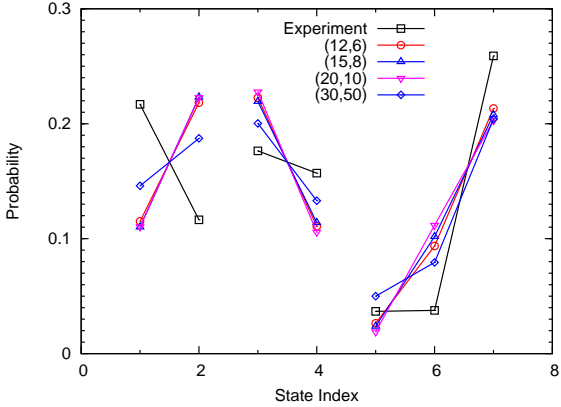}
\caption{The probabilities for the $7$ distinct mechanically 
		stable clusters of $N=4$ bubbles obtained from experiments and simulations 
		for $(m,n)=(12,6)$, $(15,8)$, $(20,10)$, and $(30,50)$ with JKR 
		mixing rules for $\epsilon_{ij}$ using protocol $1$ with $R=10^{-1}$ and 
		open boundary conditions.} 
	\label{fig:expsimN4prob}
\end{figure}

\begin{figure}
	\includegraphics[width=0.9\columnwidth]{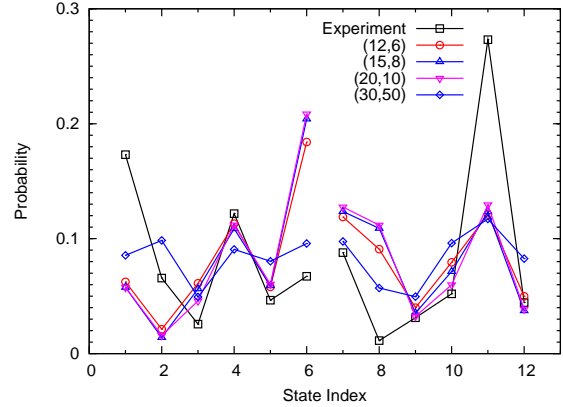}
	\caption{The probabilities for the $12$ distinct mechanically 
		stable clusters of $N=5$ bubbles obtained from experiments and simulations 
		for $(m,n)=(12,6)$, $(15,8)$, $(20,10)$, and $(30,50)$ with JKR 
		mixing rules for $\epsilon_{ij}$ using protocol $1$ with $R=10^{-1}$ under open boundaries.} 
	\label{fig:expsimN5prob}
\end{figure}

For $N=4$ and $5$, the probabilities of the clusters obtained from
both simulations (using protocol $1$) and experiments occur between
$\approx 0.01$ and $0.25$.  (See Figs.~\ref{fig:expsimN4prob}
and~\ref{fig:expsimN5prob}.) One observes discrepancies for a few
clusters ({\it i.e.} clusters $1$ and $2$ for $N=4$ and $1$, $6$, and
$11$ for $N=5$). However, overall there is qualitative agreement
between the probabilities obtained from the simulations and
experiments, even though there are key differences in the cluster
generation protocols.  For the experiments, small initial bubble
clusters are randomly generated and then they combine to form the
$N=4$ and $5$ mechanically stable clusters. In contrast, protocol $1$
in the simulations involves randomly placing bubbles in the simulation
cell and then cooling the system so that the bubbles form a single
cluster.

\begin{figure}
	\includegraphics[width=0.9\columnwidth]{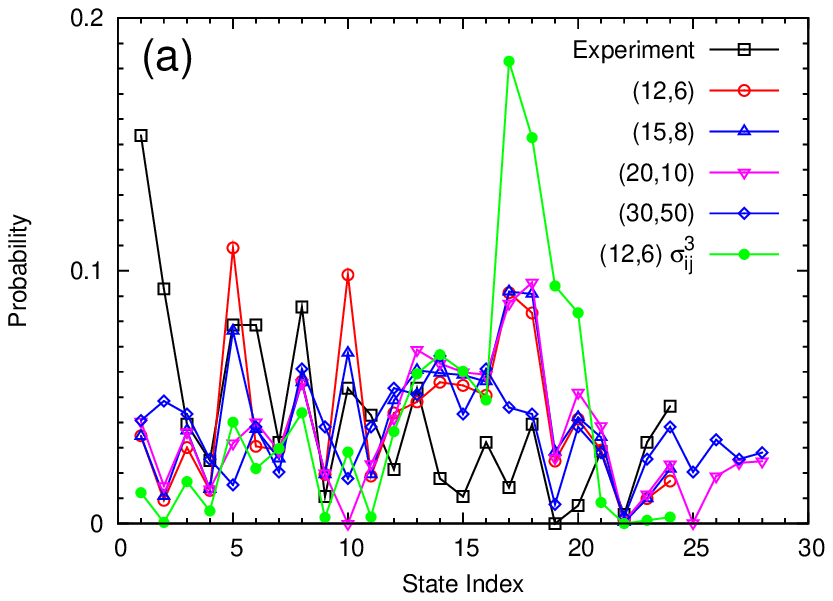}
	\includegraphics[width=0.9\columnwidth]{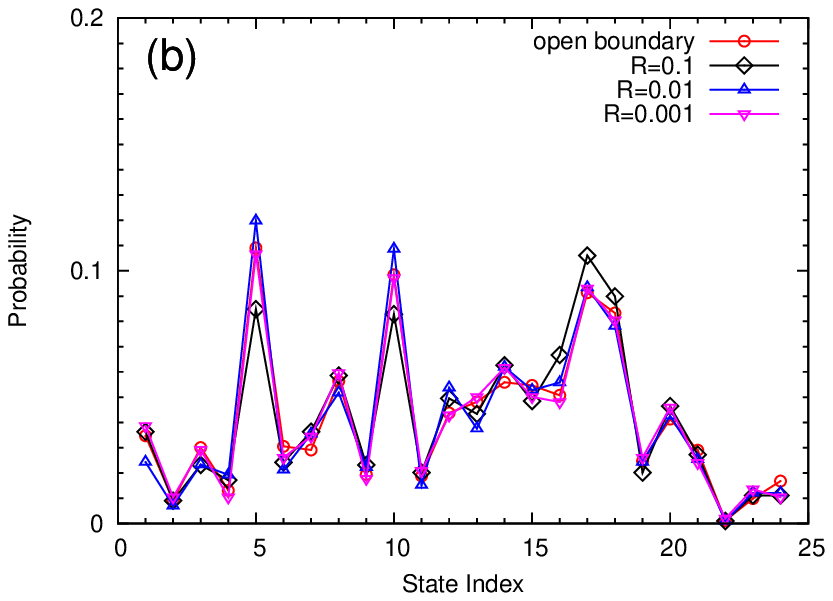}
\caption{(a) Probabilities for the mechanically stable $N=6$ clusters
(with $N_L=3$ and $N_S=3$) obtained from experiments and simulations
using protocol $1$ with cooling rate $R=10^{-1}$ for several $(m,n)$
with JKR mixing rules and the $(12,6)$ potential with $\epsilon_{ij}
\sim \sigma_{ij}^3$. In the simulations with short-ranged
attractions ({\it i.e.} $(30,50)$ and $(20,10)$), the additional
clusters $25$ and $26$ (similar to cluster $5$), and $27$ and $28$ (similar
to cluster $10$) are generated. (b) Comparison of the cluster probabilities
from simulations with the $(12,6)$ potential, JKR mixing, open
boundaries, and $R=10^{-1}$ to those with closed boundaries over a
range of cooling rates, $10^{-3} < R < 10^{-1}$.}
\label{fig:sim-expN6}
\end{figure}

For $N=4$, the cluster probabilities obtained from simulations do not
change significantly as the potential is tuned from from long- $(12,6)$
to short-ranged $(30,50)$. For $N=5$, the cluster probabilities from
the simulations become more equiprobable as the potential is tuned from
long- to short-ranged, but the effect is modest.  We find that the
changes in the cluster probabilities are even smaller when we vary the
cooling rate (using protocol $1$) over several orders of magnitude for
$N=4$ and $5$. (See Fig.~\ref{fig:sim-expN6} (b), where we show 
the effect of the cooling rate on the cluster probabilities for 
$N=6$.) Thus, studies of the cluster
probabilities for $N = 4$ and
$5$ do not allow us to determine which bubble interaction potentials and 
protocols implemented in simulations best match those in experiments. 
 
In Fig.~\ref{fig:sim-expN6} (a), we show the probabilities for the
mechanically stable $N=6$ clusters (with $N_L=3$ and $N_S=3$) from
experiments and simulations (using protocol $1$). In experiments,
clusters $1$, $2$, $5$, $6$, and $8$ are highly probable, while
clusters $19$ and $22$ are extremely rare. As discussed previously,
the simulations with long-range attractions ({\it i.e.} $(12,6)$ and
$(15,8)$) do not generate clusters $25$-$28$, whereas the simulations
with short-range attractions ({\it i.e.} $(20,10)$ and $(30,50)$) do.
For the simulations with JKR mixing rules,
most clusters have probabilities between $0.01$ and $0.1$. The cluster
probabilities from the simulations with the $(12,6)$ potential and
$\epsilon_{ij} \sim \sigma_{ij}^3$ (using protocol $1$) possess a
strong peak for clusters $13$-$20$, with smaller probabilities for
other clusters.  Even though the gap distance in cluster $5$ generated
using the $(12,6)$ potential with $\epsilon_{ij} \sim \sigma_{ij}^3$
matched that from the experiments, the probability for cluster $5$ (as
well as the probabilities for other clusters) obtained from these
simulations are not the same as those from the experiments.  These
results suggest that the cluster generation protocol can strongly
affect the cluster probabilities.

\begin{figure}
	\includegraphics[width=0.9\columnwidth]{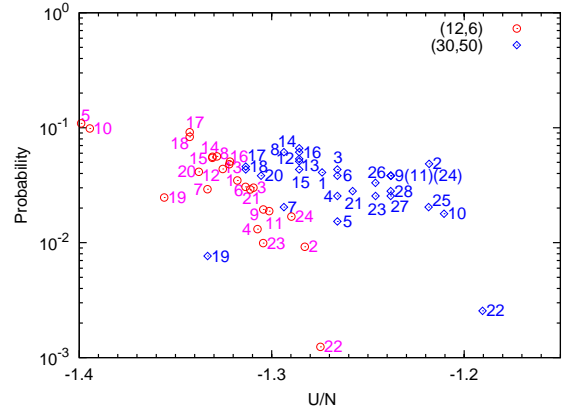}
\caption{The probability to obtain a given mechanically stable 
bubble cluster plotted versus the total potential energy per bubble $U/N$ 
from simulations using the $(12,6)$ and $(30,50)$ interaction potentials, 
JKR mixing rules, protocol $1$, and cooling rate $R=10^{-1}$.  The cluster
indexes are labelled near the data points.} 
\label{fig:prob-energy}
\end{figure}

To better understand the distinct bubble cluster probabilities, we
first investigated whether they are exponentially distributed in
energy, or otherwise strongly correlated with the energy of the
cluster. In Fig.~\ref{fig:prob-energy}, we plot the frequency with
which each distinct cluster occurs versus the total potential energy
per bubble $U/N$ of the cluster for simulations using the $(12,6)$ and
$(30,5)$ interaction potentials, JKR mixing rules, protocol $1$, and
cooling rate $R=10^{-1}$.  For the case of the $(30,50)$ potential,
there is very little correlation between probability and energy. For
the case of the $(12,6)$ potential, the probability decreases with
increasing energy, but there are large fluctuations. Given the weak
correlation between probability and energy, we now focus on the role
of the cluster formation protocol in determining the frequency with
which each cluster occurs.

As shown in Fig.~\ref{fig:sim-expN6} (a), $N=6$ clusters $19$ and $22$ are
extremely rare for experimental protocol $A$ (with mixing region
$1.9\ {\rm cm} \times 1.4\ {\rm cm}$), but they possess finite
probabilities in the simulations.  However, when we construct clusters
$19$ and $22$ one bubble at a time in the experiments, we find that
they are stable over long times.  In addition, for experimental
protocol $A$, clusters $1$ and $2$ are the most probable with
probabilities $\gtrsim 0.1$, and clusters $14$-$22$ are rare. In
contrast, using protocol $1$ from the simulations, clusters $1$ and
$2$ possess probabilities $\lesssim 0.05$ and show peaks for other
clusters that are not particularly frequent in experiments.  We show 
in Fig.~\ref{fig:sim-expN6} (b) that the differences between the 
cluster probabilities generated from experimental protocol $A$ and 
simulation protocol $1$ are not caused by cooling rate effects 
in the simulations.   

\begin{figure}
	\includegraphics[width=0.9\columnwidth]{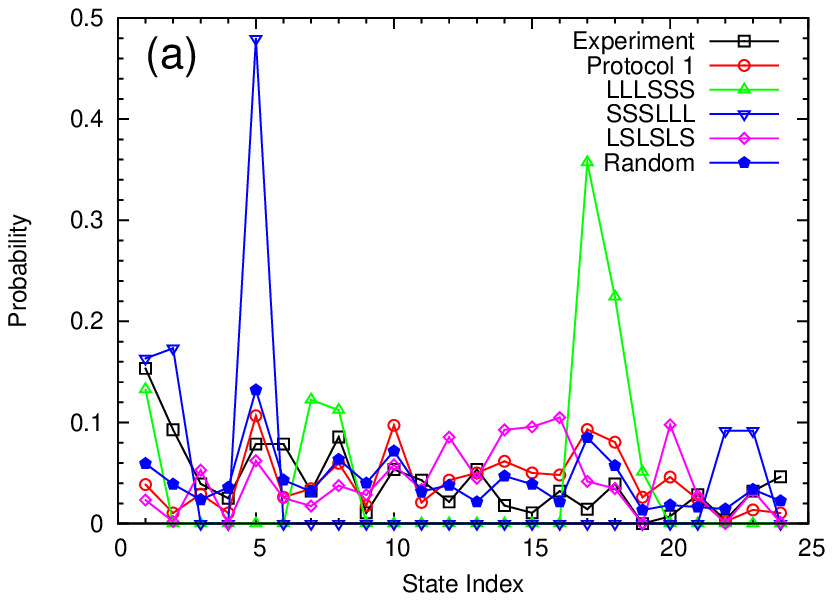}
	\includegraphics[width=0.9\columnwidth]{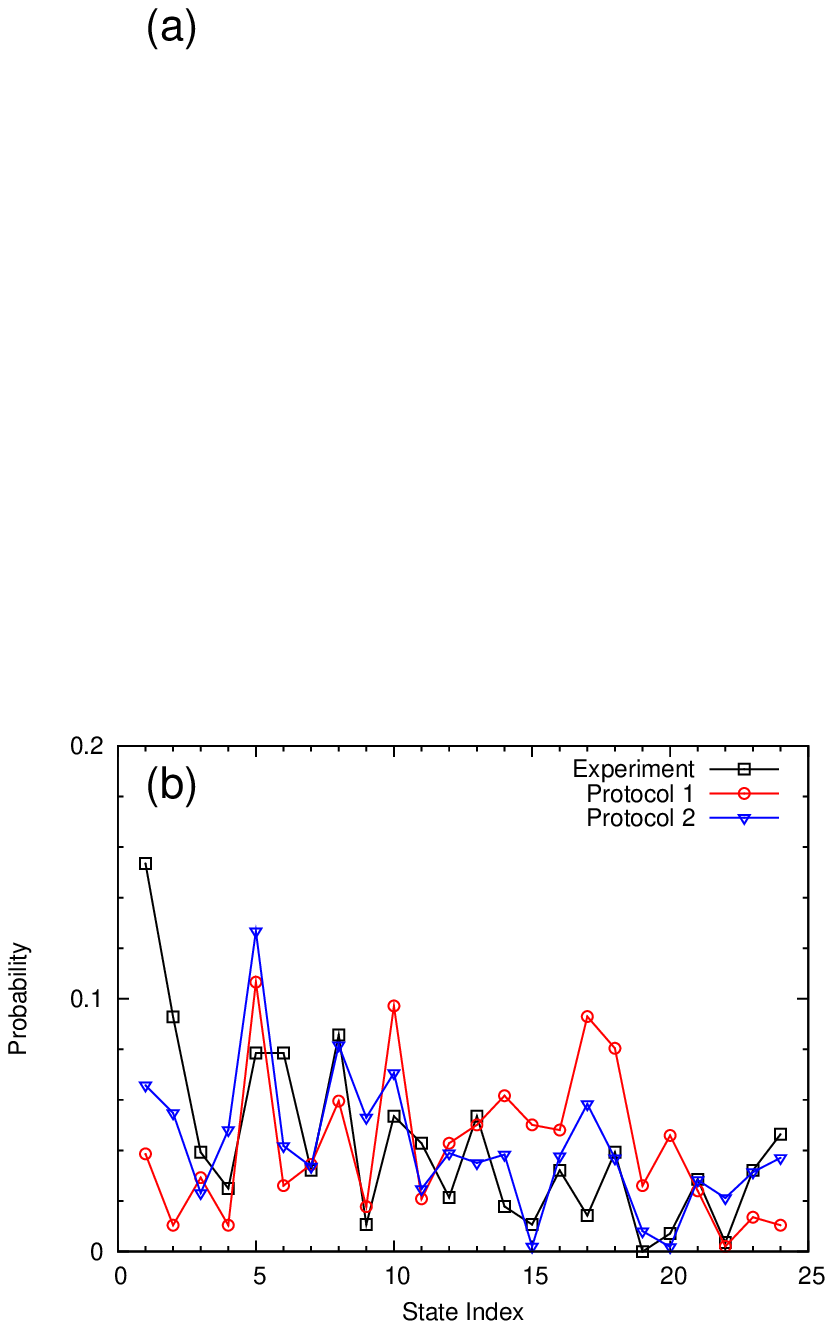}
	\caption{(a) Comparison of the $N=6$ cluster probabilities (with
		$N_L=3$ and $N_S=3$) obtained from simulations using the $(12,6)$
		potential, JKR mixing rules, cooling rate $R=10^{-3}$, and protocol
		$1$ to those obtained using protocol $2$ for several bubble
		addition sequences (LLLSSS, SSSLLL, LSLSLS, and random selection of
		small and large bubbles). (b) Same comparison of the cluster
		probabilities in (a), except for protocol $2$, we have chosen
		optimal weightings of the insertion sequences so that the
		root-mean-square deviation between the simulation and experimental
		probabilities is minimum.}
	\label{fig:rate}
\end{figure}

In Fig.~\ref{fig:rate} (a), we show the probabilities from the
simulations (using the $(12,6)$ potential and JKR mixing rules) for
protocol $2$ and several of the $20$ possible bubble addition
sequences.  Sequences SSSLLL and LLLSSS match the experimental
probability for cluster $1$, but possess strong probability peaks for
clusters $5$ and $17$, which do not occur in experiments
(with protocol $A$).  We also identified the optimal weightings of the
$20$ bubble addition sequences that provided the smallest
root-mean-square (rms) deviation between the simulation and
experimental probabilities (Fig.~\ref{fig:rate} (b)).  We find that
protocol $2$ from the simulations with optimal weights has an rms
deviation in probability from experimental protocol $A$ of $\approx 0.02$.

\begin{figure}
	\includegraphics[width=0.9\columnwidth]{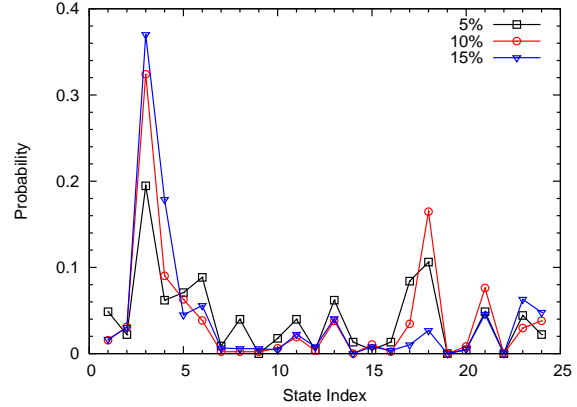}
	\caption{Probabilities for the $24$ distinct mechanically stable 
		clusters for $N=6$ bubbles (with $N_L=3$ and $N_S=3$) obtained from 
		experiments using protocol $B$ for three glycerin concentrations
		$5\%$, $10\%$, and $15\%$.} 
	\label{fig:exp-protocol}
\end{figure}

We also studied the effects of changing the experimental protocol
(from protocol $A$ to $B$ with a smaller cluster mixing region $1.2\ {\rm cm}
\times 1.4\ {\rm cm}$). As shown in Fig.~\ref{fig:exp-protocol}, for
experimental protocol $B$, cluster $1$ is rare, and clusters $3$, $4$,
$17$, and $18$ are frequent.  Variation in the glycerin concentration
from $5\%$ to $20\%$ gives rise to smaller changes in the cluster
probabilities than those caused by changes in the size of the mixing
region. These results show that for bubble rafts the way in which the
smaller clusters (with $N<6$) combine to form the $N=6$ clusters
strongly influences their probabilities.

\section{Discussion}
\label{disc}

Motivated by the potential role that small clusters of particles can
play in the mechanical response of large amorphous systems, we
systematically studied the impact of the form of the bubble
interaction potential on the formation of clusters of four, five, and
six bubbles using coordinated simulations and experiments of
two-dimensional bubble rafts. This work highlights several interesting
features of small bubble clusters. First, for four and five bubble
clusters, we find that the form of the attractive bubble interactions
does not significantly impact the structure of mechanically stable
clusters and the probability with which they occur. However, for six
bubble clusters and larger, the energy landscape is sufficiently
complex that the range and strength of the bubble interaction
potential has a significant impact on the structure of the observed
clusters.

A key result of this work is the identification of the important role
of {\it both} the range and strength of the potential in determining
the structure of the mechanically stable clusters. First, we found
that for $N \ge 6$, as the range of the bubble interaction potential
decreases from that for the $(12,6)$ potential to that for the
short-range ``sticky'' $(30,50)$ potential, the clusters tend to
possess more gaps between bubble pairs on the periphery of the
clusters.  In this case, the distinct clusters obtained from simulations
using the long-range potential are a better match to those from
experiments. In addition, changing the strength of the interaction
potential produced significant changes in the relative distances
between bubbles on the periphery in specific clusters. By increasing
the strength of bubble interaction from the JKR to $\epsilon_{ij} \sim
\sigma_{ij}^3$ mixing rules, we obtained better quantitative agreement
for the structure of the clusters from simulations and experiments.

\begin{figure}
	\begin{center}
		\includegraphics[width=0.9\columnwidth]{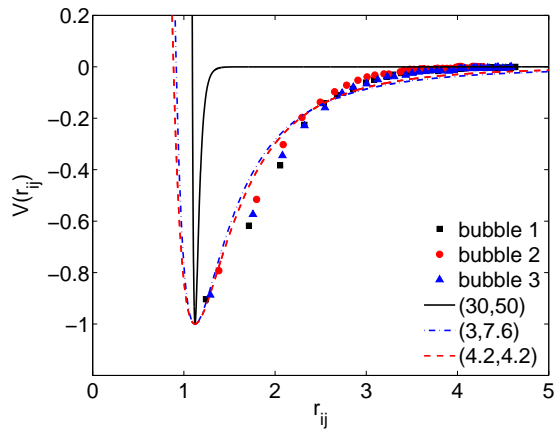}
	\end{center}
\caption{The effective potential $V(r_{ij})$ (scaled by the minimum
value) as a function of the bubble separation $r_{ij}$ for three
separate experimental trials (filled symbols). We compare the
effective potential from experiments to the generalized
Lennard-Jones pair potential with different values of $(m,n)$ (solid, 
dashed, and dot-dashed lines).}
\label{fig:bub-exp-pot}
\end{figure}

Given the importance of the form of the bubble interaction potential
in determining the structure of the bubble clusters, we carried out
preliminary measurements of the long-range attraction between bubbles
that are not in contact in the bubble raft system. As discussed in
Sec.~\ref{sec:Introduction}, there are two contributions to the
attractive force between bubbles.  One is an adhesive force between
two contacting bubbles. The second is a longer-ranged attractive force
due to distortion of the water surface by the bubbles.  In
Fig.~\ref{fig:bub-exp-pot}, we show measurements of the inter-bubble
effective potential as a function of separation and compare these results to
the pair force from the generalized Lennard-Jones potential.  The best
fits to the experimental data give $(m,n) \approx (3,8)$ or $(4,4)$,
which are much longer-ranged than $(30,50)$, depending on whether we
fix the minimum in the potential for the fit or not.  Note that the
effective potential includes contributions from drag forces in
addition to the meniscus-distortion force that are not relevant for
static clusters.

This work also investigated the protocol dependence of the frequency
with which mechanically stable bubble clusters occur. As shown in
Fig.~\ref{fig:rate} (b), we obtained the best agreement between the
cluster probabilities from experiments and simulations using protocol
$2$. However, quantitative agreement requires a detailed understanding
of how each bubble cluster is generated, {\it i.e.} are clusters
generated by adding bubbles one at a time or do smaller mechanically
stable clusters with different sizes combine to form each cluster?

The fact that the probabilities of the mechanically stable clusters
are strongly protocol dependent has important consequences for
materials design.  For example, one can change the protocol so that
clusters with particular properties are more probable.  In this
direction, we have developed protocols in experiments that allow us to
vary the cluster probabilities over a wide range, and in a controlled
fashion. A current challenge is that the experimental protocols
generate initial clusters with a distribution of sizes, whereas the
simulation protocols use particular insertion sequences. A full
exploration of different classes of cluster generation protocols will
be the focus of our future work in this area.

Though we performed preliminary studies of clusters with $N = 7$, we
only described representative clusters with the aim of confirming the
role of the form of the bubble interaction potential in determining
the structure of small clusters. We showed the impact of both the
range and strength of the potential, but the initial studies also
point to the challenges associated with enumerating the large number
of distinct clusters for increasing $N$. In future work, we will
consider larger numbers of bubbles $6 < N \le 20$ and enumerate all
mechanically stable clusters in experiments and simulations using
advanced sampling techniques~\cite{martiniani}.  These studies will
allow us to further refine the form of the bubble-bubble interaction
potential.  For each distinct mechanically stable cluster obtained in
the experiments, we can find the cluster from simulations (with a
given interaction potential) that is closest to it in configuration
space.  In this way, we can identify corresponding mechanically stable
clusters in the simulations and experiments and determine the interaction 
potential that gives the closest match~\cite{gao3}.

\section*{Acknowledgments}
\label{ack}

The authors acknowledge financial support from NSF MRSEC DMR-1119826 
(K.Z.), from NSF Grant No. CMMI-1462489 (C.O.), and from
NSF Grant DMR-1506991 (M.D.,C.K.,N.S.).  This work was 
supported by the High Performance Computing facilities operated 
by, and the staff of, the Yale Center for Research Computing.

\footnotesize{

\providecommand*{\mcitethebibliography}{\thebibliography}
\csname @ifundefined\endcsname{endmcitethebibliography}
{\let\endmcitethebibliography\endthebibliography}{}
\begin{mcitethebibliography}{57}
\providecommand*{\natexlab}[1]{#1}
\providecommand*{\mciteSetBstSublistMode}[1]{}
\providecommand*{\mciteSetBstMaxWidthForm}[2]{}
\providecommand*{\mciteBstWouldAddEndPuncttrue}
  {\def\EndOfBibitem{\unskip.}}
\providecommand*{\mciteBstWouldAddEndPunctfalse}
  {\let\EndOfBibitem\relax}
\providecommand*{\mciteSetBstMidEndSepPunct}[3]{}
\providecommand*{\mciteSetBstSublistLabelBeginEnd}[3]{}
\providecommand*{\EndOfBibitem}{}
\mciteSetBstSublistMode{f}
\mciteSetBstMaxWidthForm{subitem}
{(\emph{\alph{mcitesubitemcount}})}
\mciteSetBstSublistLabelBeginEnd{\mcitemaxwidthsubitemform\space}
{\relax}{\relax}

\bibitem[Sandfeld \emph{et~al.}(2015)Sandfeld, Budrikis, Zapperi, and
  Castellanos]{sandfeld}
S.~Sandfeld, Z.~Budrikis, S.~Zapperi and D.~F. Castellanos, \emph{Journal of
  Statistical Mechanics: Theory and Experiment}, 2015,  02011\relax
\mciteBstWouldAddEndPuncttrue
\mciteSetBstMidEndSepPunct{\mcitedefaultmidpunct}
{\mcitedefaultendpunct}{\mcitedefaultseppunct}\relax
\EndOfBibitem
\bibitem[Greer(2009)]{greer2009metallic}
A.~L. Greer, \emph{Materials Today}, 2009, \textbf{12}, 14--22\relax
\mciteBstWouldAddEndPuncttrue
\mciteSetBstMidEndSepPunct{\mcitedefaultmidpunct}
{\mcitedefaultendpunct}{\mcitedefaultseppunct}\relax
\EndOfBibitem
\bibitem[Kingery(1960)]{kingery1960}
W.~D. Kingery, \emph{Introduction to Ceramics}, John Wiley \& Sons, 1960\relax
\mciteBstWouldAddEndPuncttrue
\mciteSetBstMidEndSepPunct{\mcitedefaultmidpunct}
{\mcitedefaultendpunct}{\mcitedefaultseppunct}\relax
\EndOfBibitem
\bibitem[Schall \emph{et~al.}(2007)Schall, Weitz, and Spaepen]{SWS07}
P.~Schall, D.~A. Weitz and F.~Spaepen, \emph{Science}, 2007, \textbf{318},
  1895\relax
\mciteBstWouldAddEndPuncttrue
\mciteSetBstMidEndSepPunct{\mcitedefaultmidpunct}
{\mcitedefaultendpunct}{\mcitedefaultseppunct}\relax
\EndOfBibitem
\bibitem[Chen \emph{et~al.}(2010)Chen, Semwogerere, Sato, Breedveld, and
  Weeks]{CSSBW10}
D.~Chen, D.~Semwogerere, J.~Sato, V.~Breedveld and E.~R. Weeks, \emph{Phys.
  Rev. E}, 2010, \textbf{81}, 011403\relax
\mciteBstWouldAddEndPuncttrue
\mciteSetBstMidEndSepPunct{\mcitedefaultmidpunct}
{\mcitedefaultendpunct}{\mcitedefaultseppunct}\relax
\EndOfBibitem
\bibitem[Debrégeas \emph{et~al.}(2001)Debrégeas, Tabuteau, and
  di~Meglio]{DTM01}
G.~Debrégeas, H.~Tabuteau and J.-M. di~Meglio, \emph{Phys. Rev. Lett.}, 2001,
  \textbf{87}, 178305\relax
\mciteBstWouldAddEndPuncttrue
\mciteSetBstMidEndSepPunct{\mcitedefaultmidpunct}
{\mcitedefaultendpunct}{\mcitedefaultseppunct}\relax
\EndOfBibitem
\bibitem[Dennin(2004)]{D04}
M.~Dennin, \emph{Phys. Rev. E}, 2004, \textbf{70}, 041406\relax
\mciteBstWouldAddEndPuncttrue
\mciteSetBstMidEndSepPunct{\mcitedefaultmidpunct}
{\mcitedefaultendpunct}{\mcitedefaultseppunct}\relax
\EndOfBibitem
\bibitem[Lauridsen \emph{et~al.}(2004)Lauridsen, Chanan, and Dennin]{LCD04}
J.~Lauridsen, G.~Chanan and M.~Dennin, \emph{Phys. Rev. Lett.}, 2004,
  \textbf{93}, 018303\relax
\mciteBstWouldAddEndPuncttrue
\mciteSetBstMidEndSepPunct{\mcitedefaultmidpunct}
{\mcitedefaultendpunct}{\mcitedefaultseppunct}\relax
\EndOfBibitem
\bibitem[Lundberg \emph{et~al.}(2008)Lundberg, Krishan, Xu, O'Hern, and
  Dennin]{LKXO08}
M.~Lundberg, K.~Krishan, N.~Xu, C.~S. O'Hern and M.~Dennin, \emph{Phys. Rev.
  E}, 2008, \textbf{77}, 041505\relax
\mciteBstWouldAddEndPuncttrue
\mciteSetBstMidEndSepPunct{\mcitedefaultmidpunct}
{\mcitedefaultendpunct}{\mcitedefaultseppunct}\relax
\EndOfBibitem
\bibitem[H\'{e}braud \emph{et~al.}(1997)H\'{e}braud, Lequeux, Munch, and
  Pine]{pine}
P.~H\'{e}braud, F.~Lequeux, J.~P. Munch and D.~J. Pine, \emph{Phys. Rev.
  Lett.}, 1997, \textbf{78}, 4657\relax
\mciteBstWouldAddEndPuncttrue
\mciteSetBstMidEndSepPunct{\mcitedefaultmidpunct}
{\mcitedefaultendpunct}{\mcitedefaultseppunct}\relax
\EndOfBibitem
\bibitem[Utter and Behringer(2008)]{UB08}
B.~Utter and R.~P. Behringer, \emph{Phys. Rev. Lett.}, 2008, \textbf{100},
  208302\relax
\mciteBstWouldAddEndPuncttrue
\mciteSetBstMidEndSepPunct{\mcitedefaultmidpunct}
{\mcitedefaultendpunct}{\mcitedefaultseppunct}\relax
\EndOfBibitem
\bibitem[Keim and Arratia(2013)]{KA13}
N.~C. Keim and P.~E. Arratia, \emph{Soft Matter}, 2013, \textbf{9}, 6222\relax
\mciteBstWouldAddEndPuncttrue
\mciteSetBstMidEndSepPunct{\mcitedefaultmidpunct}
{\mcitedefaultendpunct}{\mcitedefaultseppunct}\relax
\EndOfBibitem
\bibitem[Jensen \emph{et~al.}(2014)Jensen, Weitz, and Spaepen]{JWS14}
K.~E. Jensen, D.~A. Weitz and F.~Spaepen, \emph{Phys. Rev. E}, 2014,
  \textbf{90}, 042305\relax
\mciteBstWouldAddEndPuncttrue
\mciteSetBstMidEndSepPunct{\mcitedefaultmidpunct}
{\mcitedefaultendpunct}{\mcitedefaultseppunct}\relax
\EndOfBibitem
\bibitem[Li \emph{et~al.}(2015)Li, Rieser, Liu, Durian, and Li]{LRLD15}
W.~Li, J.~M. Rieser, A.~J. Liu, D.~J. Durian and J.~Li, \emph{Phys. Rev. E},
  2015, \textbf{75}, 062212\relax
\mciteBstWouldAddEndPuncttrue
\mciteSetBstMidEndSepPunct{\mcitedefaultmidpunct}
{\mcitedefaultendpunct}{\mcitedefaultseppunct}\relax
\EndOfBibitem
\bibitem[Lauridsen \emph{et~al.}(2002)Lauridsen, Twardos, and Dennin]{LTD02}
J.~Lauridsen, M.~Twardos and M.~Dennin, \emph{Physical Review Letters}, 2002,
  \textbf{89}, 098303\relax
\mciteBstWouldAddEndPuncttrue
\mciteSetBstMidEndSepPunct{\mcitedefaultmidpunct}
{\mcitedefaultendpunct}{\mcitedefaultseppunct}\relax
\EndOfBibitem
\bibitem[Pratt and Dennin(2003)]{PD03}
E.~Pratt and M.~Dennin, \emph{Physical Review E}, 2003, \textbf{67},
  051402\relax
\mciteBstWouldAddEndPuncttrue
\mciteSetBstMidEndSepPunct{\mcitedefaultmidpunct}
{\mcitedefaultendpunct}{\mcitedefaultseppunct}\relax
\EndOfBibitem
\bibitem[Picard \emph{et~al.}(2005)Picard, and. F~Lequeux, and Bocquet]{PALB05}
G.~Picard, A.~A. and. F~Lequeux and L.~Bocquet, \emph{Physical Review E}, 2005,
  \textbf{71}, 010501\relax
\mciteBstWouldAddEndPuncttrue
\mciteSetBstMidEndSepPunct{\mcitedefaultmidpunct}
{\mcitedefaultendpunct}{\mcitedefaultseppunct}\relax
\EndOfBibitem
\bibitem[Desmond and Weeks(2015)]{DW15}
K.~W. Desmond and E.~R. Weeks, \emph{Phys. Rev. Lett.}, 2015, \textbf{115},
  098302\relax
\mciteBstWouldAddEndPuncttrue
\mciteSetBstMidEndSepPunct{\mcitedefaultmidpunct}
{\mcitedefaultendpunct}{\mcitedefaultseppunct}\relax
\EndOfBibitem
\bibitem[Fan \emph{et~al.}(2017)Fan, Wang, Zhang, Liu, Schroers, Shattuck, and
  O'Hern]{meng}
M.~Fan, M.~Wang, K.~Zhang, Y.~Liu, J.~Schroers, M.~D. Shattuck and C.~S.
  O'Hern, \emph{Phys. Rev. E}, 2017, \textbf{95}, 022611\relax
\mciteBstWouldAddEndPuncttrue
\mciteSetBstMidEndSepPunct{\mcitedefaultmidpunct}
{\mcitedefaultendpunct}{\mcitedefaultseppunct}\relax
\EndOfBibitem
\bibitem[Jiang \emph{et~al.}(1999)Jiang, Swart, Saxena, and Asipauskas]{JSSA99}
Y.~Jiang, P.~J. Swart, A.~Saxena and J.~A. Asipauskas, M.and~Glazier,
  \emph{Phys. Rev. E}, 1999, \textbf{59}, 5819\relax
\mciteBstWouldAddEndPuncttrue
\mciteSetBstMidEndSepPunct{\mcitedefaultmidpunct}
{\mcitedefaultendpunct}{\mcitedefaultseppunct}\relax
\EndOfBibitem
\bibitem[Kabla and Debrégeas(2003)]{KD03}
A.~Kabla and G.~Debrégeas, \emph{Phys. Rev. Lett.}, 2003, \textbf{90},
  258303\relax
\mciteBstWouldAddEndPuncttrue
\mciteSetBstMidEndSepPunct{\mcitedefaultmidpunct}
{\mcitedefaultendpunct}{\mcitedefaultseppunct}\relax
\EndOfBibitem
\bibitem[Wang \emph{et~al.}(2006)Wang, Krishan, and Dennin]{WKD06a}
Y.~Wang, K.~Krishan and M.~Dennin, \emph{Phys. Rev. E}, 2006, \textbf{73},
  031401\relax
\mciteBstWouldAddEndPuncttrue
\mciteSetBstMidEndSepPunct{\mcitedefaultmidpunct}
{\mcitedefaultendpunct}{\mcitedefaultseppunct}\relax
\EndOfBibitem
\bibitem[Kabla \emph{et~al.}(2007)Kabla, Scheibert, and Debregeas]{KSD07}
A.~Kabla, J.~Scheibert and G.~Debregeas, \emph{Physical Review E}, 2007,
  \textbf{587}, 45\relax
\mciteBstWouldAddEndPuncttrue
\mciteSetBstMidEndSepPunct{\mcitedefaultmidpunct}
{\mcitedefaultendpunct}{\mcitedefaultseppunct}\relax
\EndOfBibitem
\bibitem[Goyon \emph{et~al.}(2008)Goyon, Colin, Ovarlez, Ajdari, and
  Bocquet]{GCOA08}
J.~Goyon, A.~Colin, G.~Ovarlez, A.~Ajdari and L.~Bocquet, \emph{Nature}, 2008,
  \textbf{454}, 84\relax
\mciteBstWouldAddEndPuncttrue
\mciteSetBstMidEndSepPunct{\mcitedefaultmidpunct}
{\mcitedefaultendpunct}{\mcitedefaultseppunct}\relax
\EndOfBibitem
\bibitem[Divoux \emph{et~al.}(2013)Divoux, Grenard, and Manneville]{DGM13}
T.~Divoux, V.~Grenard and S.~Manneville, \emph{Phys. Rev. Lett.}, 2013,
  \textbf{110}, 018304\relax
\mciteBstWouldAddEndPuncttrue
\mciteSetBstMidEndSepPunct{\mcitedefaultmidpunct}
{\mcitedefaultendpunct}{\mcitedefaultseppunct}\relax
\EndOfBibitem
\bibitem[Spaepen(1977)]{S77}
F.~Spaepen, \emph{Acta Metall.}, 1977, \textbf{25}, 407\relax
\mciteBstWouldAddEndPuncttrue
\mciteSetBstMidEndSepPunct{\mcitedefaultmidpunct}
{\mcitedefaultendpunct}{\mcitedefaultseppunct}\relax
\EndOfBibitem
\bibitem[Argon(1979)]{A79}
A.~S. Argon, \emph{Acta Metall.}, 1979, \textbf{27}, 47\relax
\mciteBstWouldAddEndPuncttrue
\mciteSetBstMidEndSepPunct{\mcitedefaultmidpunct}
{\mcitedefaultendpunct}{\mcitedefaultseppunct}\relax
\EndOfBibitem
\bibitem[Falk and Langer(1998)]{FL98}
M.~L. Falk and J.~Langer, \emph{Phys. Rev. E}, 1998, \textbf{57}, 7192\relax
\mciteBstWouldAddEndPuncttrue
\mciteSetBstMidEndSepPunct{\mcitedefaultmidpunct}
{\mcitedefaultendpunct}{\mcitedefaultseppunct}\relax
\EndOfBibitem
\bibitem[Eastgate \emph{et~al.}(2003)Eastgate, Langer, and
  Pechenik]{ELP_PRL_2003}
L.~Eastgate, J.~Langer and L.~Pechenik, \emph{Phys. Rev. Lett.}, 2003,
  \textbf{90}, 045506\relax
\mciteBstWouldAddEndPuncttrue
\mciteSetBstMidEndSepPunct{\mcitedefaultmidpunct}
{\mcitedefaultendpunct}{\mcitedefaultseppunct}\relax
\EndOfBibitem
\bibitem[Langer and Pechenik(2003)]{LP03}
J.~S. Langer and L.~Pechenik, \emph{Phys. Rev. E}, 2003, \textbf{68},
  061507\relax
\mciteBstWouldAddEndPuncttrue
\mciteSetBstMidEndSepPunct{\mcitedefaultmidpunct}
{\mcitedefaultendpunct}{\mcitedefaultseppunct}\relax
\EndOfBibitem
\bibitem[Langer(2004)]{L04}
J.~S. Langer, \emph{Phys. Rev. E}, 2004, \textbf{70}, 041502\relax
\mciteBstWouldAddEndPuncttrue
\mciteSetBstMidEndSepPunct{\mcitedefaultmidpunct}
{\mcitedefaultendpunct}{\mcitedefaultseppunct}\relax
\EndOfBibitem
\bibitem[Pechenik(2005)]{P05}
L.~Pechenik, \emph{Phys. Rev. E}, 2005, \textbf{72}, 021507\relax
\mciteBstWouldAddEndPuncttrue
\mciteSetBstMidEndSepPunct{\mcitedefaultmidpunct}
{\mcitedefaultendpunct}{\mcitedefaultseppunct}\relax
\EndOfBibitem
\bibitem[Mayr(2006)]{M06}
S.~G. Mayr, \emph{Phys. Rev. Lett.}, 2006, \textbf{{97}}, 195501\relax
\mciteBstWouldAddEndPuncttrue
\mciteSetBstMidEndSepPunct{\mcitedefaultmidpunct}
{\mcitedefaultendpunct}{\mcitedefaultseppunct}\relax
\EndOfBibitem
\bibitem[Langer(2006)]{L06}
J.~Langer, \emph{Phys. Rev. E}, 2006, \textbf{73}, 041504\relax
\mciteBstWouldAddEndPuncttrue
\mciteSetBstMidEndSepPunct{\mcitedefaultmidpunct}
{\mcitedefaultendpunct}{\mcitedefaultseppunct}\relax
\EndOfBibitem
\bibitem[Bouchbinder \emph{et~al.}(2007)Bouchbinder, Langer, and
  Procaccia]{BLP07a}
E.~Bouchbinder, J.~Langer and I.~Procaccia, \emph{Phys. Rev. E}, 2007,
  \textbf{75}, 036107\relax
\mciteBstWouldAddEndPuncttrue
\mciteSetBstMidEndSepPunct{\mcitedefaultmidpunct}
{\mcitedefaultendpunct}{\mcitedefaultseppunct}\relax
\EndOfBibitem
\bibitem[Bouchbinder \emph{et~al.}(2007)Bouchbinder, Langer, and
  Procaccia]{BLP07b}
E.~Bouchbinder, J.~Langer and I.~Procaccia, \emph{Phys. Rev. E}, 2007,
  \textbf{75}, 036108\relax
\mciteBstWouldAddEndPuncttrue
\mciteSetBstMidEndSepPunct{\mcitedefaultmidpunct}
{\mcitedefaultendpunct}{\mcitedefaultseppunct}\relax
\EndOfBibitem
\bibitem[Langer(2008)]{L08}
J.~Langer, \emph{Phys. Rev. E}, 2008, \textbf{77}, 021502\relax
\mciteBstWouldAddEndPuncttrue
\mciteSetBstMidEndSepPunct{\mcitedefaultmidpunct}
{\mcitedefaultendpunct}{\mcitedefaultseppunct}\relax
\EndOfBibitem
\bibitem[Manning and Liu(2011)]{ML11}
M.~L. Manning and A.~J. Liu, \emph{Phys. Rev. Lett.}, 2011, \textbf{107},
  108302\relax
\mciteBstWouldAddEndPuncttrue
\mciteSetBstMidEndSepPunct{\mcitedefaultmidpunct}
{\mcitedefaultendpunct}{\mcitedefaultseppunct}\relax
\EndOfBibitem
\bibitem[Schoenholz \emph{et~al.}(2014)Schoenholz, Liu, Riggleman, and
  Rottler]{SLRR14}
S.~S. Schoenholz, A.~Liu, R.~Riggleman and J.~Rottler, \emph{Phys. Rev. X},
  2014, \textbf{4}, 031014\relax
\mciteBstWouldAddEndPuncttrue
\mciteSetBstMidEndSepPunct{\mcitedefaultmidpunct}
{\mcitedefaultendpunct}{\mcitedefaultseppunct}\relax
\EndOfBibitem
\bibitem[Tong and Xu(2014)]{tong}
H.~Tong and N.~Xu, \emph{Phys. Rev. E}, 2014, \textbf{90}, 010401(R)\relax
\mciteBstWouldAddEndPuncttrue
\mciteSetBstMidEndSepPunct{\mcitedefaultmidpunct}
{\mcitedefaultendpunct}{\mcitedefaultseppunct}\relax
\EndOfBibitem
\bibitem[Pouliquen and Forterre(2009)]{PF09}
O.~Pouliquen and Y.~Forterre, \emph{Phil. trans. R. Soc. A}, 2009,
  \textbf{367}, 5091\relax
\mciteBstWouldAddEndPuncttrue
\mciteSetBstMidEndSepPunct{\mcitedefaultmidpunct}
{\mcitedefaultendpunct}{\mcitedefaultseppunct}\relax
\EndOfBibitem
\bibitem[Kamrin and Koval(2012)]{KK12}
K.~Kamrin and G.~Koval, \emph{Phys. Rev. Lett.}, 2012, \textbf{108},
  178301\relax
\mciteBstWouldAddEndPuncttrue
\mciteSetBstMidEndSepPunct{\mcitedefaultmidpunct}
{\mcitedefaultendpunct}{\mcitedefaultseppunct}\relax
\EndOfBibitem
\bibitem[Tewari \emph{et~al.}(1999)Tewari, Schiemann, Durian, Knobler, Langer,
  and Liu]{TSDK99}
S.~Tewari, D.~Schiemann, D.~J. Durian, C.~M. Knobler, S.~A. Langer and A.~J.
  Liu, \emph{Phys. Rev. E}, 1999, \textbf{60}, 4385\relax
\mciteBstWouldAddEndPuncttrue
\mciteSetBstMidEndSepPunct{\mcitedefaultmidpunct}
{\mcitedefaultendpunct}{\mcitedefaultseppunct}\relax
\EndOfBibitem
\bibitem[Durand and Stone(2006)]{DS06}
M.~Durand and H.~A. Stone, \emph{Phys. Rev. Lett.}, 2006, \textbf{97},
  226101\relax
\mciteBstWouldAddEndPuncttrue
\mciteSetBstMidEndSepPunct{\mcitedefaultmidpunct}
{\mcitedefaultendpunct}{\mcitedefaultseppunct}\relax
\EndOfBibitem
\bibitem[Vaz and Fortes(2001)]{VF01}
M.~F. Vaz and M.~A. Fortes, \emph{Journal of Physics: Condensed Matter}, 2001,
  \textbf{13}, 1395\relax
\mciteBstWouldAddEndPuncttrue
\mciteSetBstMidEndSepPunct{\mcitedefaultmidpunct}
{\mcitedefaultendpunct}{\mcitedefaultseppunct}\relax
\EndOfBibitem
\bibitem[Shen \emph{et~al.}(2014)Shen, Papanikolaou, O'Hern, and
  Shattuck]{shen}
T.~Shen, S.~Papanikolaou, C.~S. O'Hern and M.~D. Shattuck, \emph{Phys. Rev.
  Lett.}, 2014, \textbf{113}, 128302\relax
\mciteBstWouldAddEndPuncttrue
\mciteSetBstMidEndSepPunct{\mcitedefaultmidpunct}
{\mcitedefaultendpunct}{\mcitedefaultseppunct}\relax
\EndOfBibitem
\bibitem[Kralchevsky and Nagayama(2000)]{KN00}
P.~A. Kralchevsky and K.~Nagayama, \emph{Advances in Colloid and Interface
  Science}, 2000, \textbf{85}, 145 -- 192\relax
\mciteBstWouldAddEndPuncttrue
\mciteSetBstMidEndSepPunct{\mcitedefaultmidpunct}
{\mcitedefaultendpunct}{\mcitedefaultseppunct}\relax
\EndOfBibitem
\bibitem[Kralchevsky and Denkov(2001)]{KD01}
P.~A. Kralchevsky and N.~D. Denkov, \emph{Curr. Opin. Colloid Interface Sci.},
  2001, \textbf{6}, 383\relax
\mciteBstWouldAddEndPuncttrue
\mciteSetBstMidEndSepPunct{\mcitedefaultmidpunct}
{\mcitedefaultendpunct}{\mcitedefaultseppunct}\relax
\EndOfBibitem
\bibitem[Bragg(1942)]{B42}
L.~Bragg, \emph{Journal of Scientific Instruments}, 1942, \textbf{19},
  148\relax
\mciteBstWouldAddEndPuncttrue
\mciteSetBstMidEndSepPunct{\mcitedefaultmidpunct}
{\mcitedefaultendpunct}{\mcitedefaultseppunct}\relax
\EndOfBibitem
\bibitem[Bragg and Nye(1947)]{BN47}
L.~Bragg and J.~F. Nye, \emph{Proceedings of the Royal Society of London,
  Series A}, 1947, \textbf{190}, 474\relax
\mciteBstWouldAddEndPuncttrue
\mciteSetBstMidEndSepPunct{\mcitedefaultmidpunct}
{\mcitedefaultendpunct}{\mcitedefaultseppunct}\relax
\EndOfBibitem
\bibitem[Durian(1997)]{durian}
D.~J. Durian, \emph{Phys. Rev. E}, 1997, \textbf{55}, 1739\relax
\mciteBstWouldAddEndPuncttrue
\mciteSetBstMidEndSepPunct{\mcitedefaultmidpunct}
{\mcitedefaultendpunct}{\mcitedefaultseppunct}\relax
\EndOfBibitem
\bibitem[Johnson \emph{et~al.}(1971)Johnson, Kendall, and
  Roberts]{johnson:1971}
K.~Johnson, K.~Kendall and A.~Roberts, Proceedings of the Royal Society of
  London A: Mathematical, Physical and Engineering Sciences, 1971, pp.
  301--313\relax
\mciteBstWouldAddEndPuncttrue
\mciteSetBstMidEndSepPunct{\mcitedefaultmidpunct}
{\mcitedefaultendpunct}{\mcitedefaultseppunct}\relax
\EndOfBibitem
\bibitem[Israelachvili(2011)]{israelachvili:2011}
J.~N. Israelachvili, \emph{Intermolecular and Surface Forces}, Academic press,
  2011\relax
\mciteBstWouldAddEndPuncttrue
\mciteSetBstMidEndSepPunct{\mcitedefaultmidpunct}
{\mcitedefaultendpunct}{\mcitedefaultseppunct}\relax
\EndOfBibitem
\bibitem[Toxvaerd and Dyre(2011)]{dyre}
S.~Toxvaerd and J.~C. Dyre, \emph{J. Chem. Phys.}, 2011, \textbf{134},
  081102\relax
\mciteBstWouldAddEndPuncttrue
\mciteSetBstMidEndSepPunct{\mcitedefaultmidpunct}
{\mcitedefaultendpunct}{\mcitedefaultseppunct}\relax
\EndOfBibitem
\bibitem[Tkachenko and Witten(1999)]{witten}
A.~V. Tkachenko and T.~A. Witten, \emph{Phys. Rev. E}, 1999, \textbf{60},
  687\relax
\mciteBstWouldAddEndPuncttrue
\mciteSetBstMidEndSepPunct{\mcitedefaultmidpunct}
{\mcitedefaultendpunct}{\mcitedefaultseppunct}\relax
\EndOfBibitem
\bibitem[Martiniani \emph{et~al.}(2016)Martiniani, Schrenk, Stevenson, Wales,
  and Frenkel]{martiniani}
S.~Martiniani, K.~J. Schrenk, J.~D. Stevenson, D.~J. Wales and D.~Frenkel,
  \emph{Phys. Rev. E}, 2016, \textbf{93}, 012906\relax
\mciteBstWouldAddEndPuncttrue
\mciteSetBstMidEndSepPunct{\mcitedefaultmidpunct}
{\mcitedefaultendpunct}{\mcitedefaultseppunct}\relax
\EndOfBibitem
\bibitem[Gao \emph{et~al.}(2009)Gao, Blawzdziewicz, O'Hern, and Shattuck]{gao3}
G.-J. Gao, J.~Blawzdziewicz, C.~S. O'Hern and M.~D. Shattuck, \emph{Phys. Rev.
  E}, 2009, \textbf{80}, 061304\relax
\mciteBstWouldAddEndPuncttrue
\mciteSetBstMidEndSepPunct{\mcitedefaultmidpunct}
{\mcitedefaultendpunct}{\mcitedefaultseppunct}\relax
\EndOfBibitem
\end{mcitethebibliography}
\providecommand*{\mcitethebibliography}{\thebibliography}
\csname @ifundefined\endcsname{endmcitethebibliography}
{\let\endmcitethebibliography\endthebibliography}{}

\bibliographystyle{rsc} 
}

\end{document}